\documentclass[a4paper,twocolumn,aps,10pt,english,amsmath,amssymb,showpacs,floatfix,notitlepage,nofootinbib,superscriptaddress,longbibliography]{revtex4}
\usepackage{amssymb}
\usepackage{dcolumn}
\usepackage{amsmath}
\usepackage{bm}
\usepackage{textcomp}
\usepackage{epstopdf}
\usepackage{color}
\usepackage{ulem}

\usepackage{graphicx}
\usepackage[sort&compress]{natbib}
\usepackage{setspace}
\usepackage{mathrsfs}
\usepackage[utf8]{inputenc}
\usepackage{txfonts}
\usepackage{multirow}
\usepackage[unicode=true,pdfusetitle,
	bookmarks=true,bookmarksnumbered=false,bookmarksopen=false,
breaklinks=false,pdfborder={0 0 1},colorlinks=false]{hyperref}
\hypersetup{colorlinks,citecolor=blue,linkcolor=blue}
\begin{document}

\title{Non-critical fluctuations of (net) charges and (net) protons from {\tt iEBE-VISHNU} hybrid model}
\author{Jixing Li}
\email{jixli2014@pku.edu.cn}
\affiliation{Department of Physics and State Key Laboratory of Nuclear Physics and
Technology, Peking University, Beijing 100871, China}
\affiliation{Collaborative Innovation Center of Quantum Matter, Beijing 100871, China}
\author{Hao-jie Xu}
\email{haojiexu@zjhu.edu.cn}
\affiliation{School of Science, Huzhou University, Huzhou 313000, China}
\affiliation{Department of Physics and State Key Laboratory of Nuclear Physics and
Technology, Peking University, Beijing 100871, China}
\author{Huichao Song}
\email{huichaosong@pku.edu.cn}
\affiliation{Department of Physics and State Key Laboratory of Nuclear Physics and
Technology, Peking University, Beijing 100871, China}
\affiliation{Collaborative Innovation Center of Quantum Matter, Beijing 100871, China}
\affiliation{Center for High Energy Physics, Peking University, Beijing 100871, China}

\date{\today}

%\begin{spacing}{2.0}

\begin{abstract}	
 In this paper, we investigate the non-critical fluctuations of (net) charges
  and (net) protons in Au+Au collisions at $\sqrt{s_{NN}}$ = 7.7, 39 and 200 GeV,
	using {\tt{iEBE-VISHNU}} hybrid model with Poisson fluctuations added in the particle event
    generator between hydrodynamics and {\tt UrQMD}. Various effects, such as volume fluctuations
	hadronic  evolution and scatterings, resonance decays, as well as realistic centrality cuts and acceptance cuts
    have been embedded in our model calculations. With properly tuned parameters, {\tt{iEBE-VISHNU}} roughly describe the centrality dependent moments and cumulants of (net) charges and (net) protons measured in experiment.  Further comparison simulations show that the volume fluctuation is the dominant factor to influence the multiplicity fluctuations, which makes the higher moments of (net) charges largely deviate from the Poison baselines.
    We also find that the effects from hadronic evolutions and resonance decays are pretty small or even negligible for the multiplicity fluctuations of both (net) charges and (net) protons.
\end{abstract}


\maketitle

\section{Introduction}\label{introduction}
Exploring the QCD phase structure of the strongly interacting matter is one of the major goals of relativistic heavy-ion collisions (RHIC)~\cite{Rev-Arsene:2004fa,Aoki:2006we,Aggarwal:2010cw,Akiba:2015jwa}. As a unique feature of the phase diagram, the QCD critical point has attracted particular interests from both theoretical and experimental sides~\cite{Stephanov:2004wx, Asakawa:2015ybt, Luo:2017faz}. It is proposed that the higher moments of conserved quantities are sensitive observable to probe the QCD critical point~\cite{Stephanov:2008qz,Athanasiou:2010kw}.
The recent Beam Energy Scan (BES) program has systematically measured the higher moments (cumulants) of net charges and net protons in Au+Au collisions at
$\sqrt{s_{NN}} = 7.7$,  $11.5$,  $14.5$,  $19.6$,  $27$, $39$, $62.4$ and $200$ GeV~\cite{Aggarwal:2010wy,Adamczyk:2013dal,Adamczyk:2014fia}.
It was found that the cumulant ratio $\kappa \sigma^2$ of net protons obviously deviates from the Poisson baselines and shows a non-monantoic behavior at lower collision energies, which indicates the potential of discovery the QCD critical point in experiments~\cite{Luo:2015ewa}.

%~\cite{Chatterjee:2009km,Garg:2013ata,Karsch:2010ck, Alba:2014eba,BraunMunzinger:2011dn}~\red{the very beginning paper}.

Besides studying the critical fluctuations, it is important to systematically investigate the non-critical/thermal fluctuations of produced hadrons for the location of the QCD critical point~\cite{Asakawa:2015ybt,Luo:2017faz,Luo:2014tga,Netrakanti:2014mta,Cleymans:2004iu,Begun:2004gs,Begun:2006uu,
Chatterjee:2009km,Garg:2013ata,Karsch:2010ck,Alba:2014eba,Landau:1980mil,
Fu:2013gga,Bhattacharyya:2015zka,BraunMunzinger:2011dn,Nahrgang:2014fza,Mishra:2016qyj,
Bazavov:2012jq,Borsanyi:2013hza}.
In traditional Hadron Resonance Gas (HRG) model with Boltzmann approximations, the thermal fluctuations are governed by the Poisson statistics~\cite{Karsch:2010ck,Garg:2013ata,Landau:1980mil}. Correspondingly, the Poisson expectations are  served as the basic thermal fluctuations baselines, which have been widely used  in both experimental analysis and theoretical study~\cite{Aggarwal:2010wy,Adamczyk:2013dal,Adamczyk:2014fia,Luo:2015ewa, Luo:2014tga, Netrakanti:2014mta,Jiang:2015hri,Jiang:2017mji}. However, a realistic heavy ion collision involves many complicated processes. Many factors could make the measured multiplicity fluctuations deviate from the Poisson baselines.  For example, the system size of the collision systems within a centrality bin fluctuate event by event, the related volume fluctuations could change the cumulants of the multiplicity distributions~\cite{Jeon:2003gk,Skokov:2012ds,Xu:2016qzd,Xu:2016skm}. Meanwhile, the finite acceptance window and acceptance efficiency also bias the multiplicity distributions measured in experiments~\cite{Bzdak:2012ab,Karsch:2015zna,Petersen:2015pcy,Ling:2015yau}. In Ref.~\cite{Bzdak:2012an,Jeon:2000wg,Sakaida:2014pya}, it was found the global conservation of baryon number, strangeness number and electric number modify the cumulants of net charges and net baryons. Besides, the isospin-randomization progress~\cite{Kitazawa:2011wh,Kitazawa:2012at}, the weak decay and other related progress also influence the fluctuations of the final produced hadrons to some extend (For related review on non-critical fluctuations, please refer to~\cite{Asakawa:2015ybt, Luo:2017faz}).

Many past research of HRG model and Lattice QCD simulations assume the system is static and in global chemical and thermal equilibrium~\cite{Begun:2006uu,Chatterjee:2009km,Garg:2013ata,Karsch:2010ck,Alba:2014eba,
Bhattacharyya:2015zka,BraunMunzinger:2011dn,Mishra:2016qyj,Nahrgang:2014fza,
Bazavov:2012jq,Borsanyi:2013hza}. However, the QGP fireball created in a relativistic heavy ion collisions is a dynamically evolving system,
where the chemical and thermal equilibrium can not be maintained during the late hadronic
evolution~\cite{Hirano:2005xf,Teaney:2002aj,Hirano:2002ds,Kolb:2002ve,Huovinen:2007xh}. Within the framework of {\tt URQMD}~\cite{Xu:2016qjd}, Luo and his collaborators have systematically calculated the thermal fluctuation baselines of final produced hadrons in Au+Au collisions at $\sqrt{s_{NN}} = 7.7 \ -  \ 200$ GeV. However, their model simulations assumed that the created systems are purely hadronic at various collision energies, which neglected the collective expansion of the QGP phase at higher collision energies.

In this paper, we will investigate the multiplicity fluctuations of (net) charges and (net) protons, using {\tt iEBE-VISHNU} hybrid model that combines viscous hydrodynamics for the expansion of the QGP with a hadron cascade model for the evolution of the hadronic matter. Compared with  other dynamical model simulations, such as  {\tt UrQMD}~\cite{Xu:2016qjd,He:2017zpg,Yang:2016xga,Zhou:2017jfk} and {\tt JAM}~\cite{He:2016uei}, we input Poisson fluctuations in the particle event generator between the hydrodynamics and hadron cascade simulations. We focus on investigating how various effects, such as volume fluctuations,  hadronic scatterings, resonance decays, centrality cut and acceptance cut, etc., influence the multiplicity fluctuations of final produced hadrons. Considering that the event-by-event simulations of {\tt iEBE-VISHNU} hybrid model are time-consuming, we only perform the simulations
at three selected collision energies, $\sqrt{s_{NN}}=200$, $39$ and $7.7$ GeV,
where the net baryon density gradually increases from higher to lower collision energies. The following related calculations will show that effects from hadronic evolution are pretty small or even negligible for the multiplicity fluctuations of (net) charges and (net) protons, which may help to largely increase the numerical efficiency for the massive data simulations in the near future.

The paper is organized as the follows: Sec.~II and Sec.~III introduce {\tt{iEBE-VISHNU}} hybrid model, the observables of multiplicity fluctuations and the set-ups of calculations. In Sec.~IV. we present the calculations for the moments and moment products of (net) charges and (net)-protons in Au+Au collisions at 7.7,  39 and 200 GeV from {\tt{iEBE-VISHNU}}, together with a comparison to the STAR data. We also investigate the effects from volume fluctuations, resonance decays and
hadronic evolution for the multiplicity fluctuations of (net) charges and (net)-protons. In Sec.V, we briefly summarize this paper.

\section{The model and Setups}\label{Model_setup}
\begin{table*}
   \vspace{2mm}
\renewcommand\arraystretch{1.2}
  \caption{Parameter set-ups of {\tt iEBE-VISHNU} for Au+Au collisions at $\sqrt{s_{NN}}=7.7$, 39 and 200 GeV.
	\vspace{3mm}
   \label{parameters}}
  \begin{tabular}{  c | c c c c c c c c c c }
     %after \\: \hline or \cline{col1-col2} \cline{col3-col4} ...
    \hline
     \hline
    \multicolumn{2}{c}{$Centrality.$ }& 0-5\%~~ & 5-10\%~~ & 10-20\% ~~&20-30\% ~~&30-40\% ~~& 40-50\%~~ & 50-60\% ~~& 60-70\%~~ & 70-80\% ~~\\
    \hline
    \multirow{3}*{\text{~7.7  GeV}} & $\mu_B(\mathrm{MeV})$ & 312.0& 311.0 & 308.5 & 307.0 &306.0 & 306.0 &290.0& 260.0 & 230.0\\
    & $\mu_Q(\mathrm{MeV})$ & 2.00 & 2.00 &2.00 &2.00 &0.80 &0.80 & 0.80 &0.80 &0.80 \\
    & \multicolumn{10}{l}{$\mu_S=69.2 (\mathrm{MeV})\;\;\;~~~~~~~~~~\gamma_{S}= -0.038\;\;\;~~~~~~~~~~ T_{c}=143.2(\mathrm{MeV})$} \\
    \hline
    \multirow{3}*{\text{39 GeV}} & $\mu_B(\mathrm{MeV})$ &86.5& 85.5 & 83.0 & 80.5 &79.6 & 75.0 & 75.0& 73.5 & 73.5 \\
	& $\mu_Q(\mathrm{MeV})$ & -0.62 &-0.54 &-0.68 &-0.68 &-0.68 &-0.67 &-0.68 &-0.68 &-0.68 \\
    & \multicolumn{10}{l}{$\mu_S= 19.5(\mathrm{MeV})\;\;\;~~~~~~~~~~\gamma_{S}= -0.052\;\;\;~~~~~~~~~~ T_{c}=155.5(\mathrm{MeV})$} \\
    \hline
    \multirow{3}*{\text{~200 GeV}} & $\mu_B(\mathrm{MeV})$ & 21.6& 21.9 & 20.8 & 19.5 & 19.4 & 17.8 & 17.7& 16.1 & 16.6 \\
	& $\mu_Q(\mathrm{MeV})$ & $8e^{-3}$ & $-3e^{-3}$ & $-5e^{-3}$ & $6e^{-3}$ &$4e^{-3}$ & $6e^{-4}$ & $1e^{-3}$ &$-7e^{-3}$ &$-6e^{-3}$ \\
  & \multicolumn{10}{l}{$\mu_S= 3.2(\mathrm{MeV})\;\;\;~~~~~~~~~~~~\gamma_{S}= -0.031\;\;\; ~~~~~~~~~~T_{c}=149.9(\mathrm{MeV})$} \\
    \hline
    \hline
  \end{tabular}
  \vspace{3mm}
\end{table*}

In this paper, we investigate the non-critical multiplicity fluctuation of
(net) charges and (net) protons in Au+Au collisions at $\sqrt{s_{NN}}$ =
7.7, 39 and 200 GeV, using {\tt{iEBE-VISHNU}}
hybrid model.  {\tt{iEBE-VISHNU}}~\cite{Shen:2014vra}
is an event-by-event version of {\tt VISHNU}, which combines viscous
hydrodynamics for the QGP expansion with a hadron cascade model for the hadronic
evolution~\cite{Song:2010aq}.
It contains four main components to simulate different
stages of a relativistic heavy ion collision: (1) the initial conditions, which are generated by some initial condition models, such as Monte-Carlo Glauber model ({\tt MC-Glb})~\cite{Miller:2007ri}, Monte-Carlo KLN
model ({\tt MC-KLN})~\cite{ Drescher:2006ca,Hirano:2009ah}. {\tt{TRENTo}} model~\cite{Moreland:2014oya}, {\tt AMPT}~\cite{Xu:2016hmp}, etc. (2) the macroscopic expansion of the QGP fluid, which is simulated by a (2+1)-dimensional viscous hydrodynamics {\tt{VISH2+1}}~\cite{Song:2007fn,Song:2009gc}. (3) the switching between the hydrodynamics and the succeeding hadron cascade simulations, which is realized by a Monte-Carlo event generator that samples particles on the switching hyper-surface with the Cooper-Frye formula~\cite{Song:2010aq}. (4) The microscopic evolution and decoupling of the hadron resonance gas, which is simulated by Ultra-relativistic Quantum Molecular Dynamics ({\tt{UrQMD}}) hadron cascade model~\cite{Bass:1998ca,Bleicher:1999xi}.

In the following text, we will introduce step (3) in more details since it is directly related to the thermal fluctuations investigated in this paper. For other details of {\tt{iEBE-VISHNU}} hybrid model, please refer to~\cite{Shen:2014vra,ShenPhD, Song:2017wtw}. From the macroscopic hydrodynamics to the microscopic {\tt UrQMD} simulations, the thermal hadrons  emitted from the switching hyper-surface are sampled according to the differential Cooper-Frye formula~\cite{Cooper:1974mv,Song:2010aq}:
\begin{equation}
E\frac{d^3 N_{i}}{dp^3}(x) = \frac{g_i}{(2\pi)^3} p^{\mu} d^3 \sigma_{\mu}(x) f(x,p),
\label{eq:primordial}
\end{equation}
where $x=(\tau, \vec{x}_{\perp}, \eta_s)$, $p=(E,\vec{p}_{\perp}, y)$
are position and 4-momentum of the emitted hadrons, $d^3 \sigma_{\mu}$
is the surface element of the switching hyper-surface $\Sigma$, and $g_i$ is the spin
degeneracy of the $i_\mathrm{th}$ hadrons. The distribution function $f(x,p) = f_0 + \delta f $,
where $f_{0}$ is the equilibrium distribution function and
$\delta f = \frac{p^{\mu}p^{\nu}\pi_{\mu\nu}}{2T^2(e+p)}f_0(1 \mp f_0)$ is the corresponding
viscous correction~\cite{Song:2007fn,Song:2009gc}.  Following~\cite{Abelev:2008ab},  the equilibrium distribution function
is taken the form: $f_0 = 1 / (\gamma_s^{-|S_i|} \text{e}^{(p^{\nu} \cdot u_{\nu}-\vec{c_i}
\cdot \vec{\mu_i})/T} \pm 1)$, where $\gamma_s$  is the strangeness saturation factor and $|S_i|$ the total number of strange and anti-strange quarks of hadron species $i$. $\vec{\mu_i}=(\mu_B,\mu_S, \mu_Q)$ are the chemical potentials of net baryons, strangeness, and electric charges,  $\vec{c_i}=(B_i, Q_i, S_i)$ are the corresponding conserved charges. In many traditional hybrid model simulations~\cite{Song:2013qma,Zhu:2016puf,Bernhard:2016tnd}, $\mu_B, \ \mu_S$ and $\mu_Q$ are all set to zero. Here, these additional tunable parameters help to achieve a nice description of the mean values of positive (negative) charges and (anti)-protons at various centralities and collision energies, which are important for the investigations of multiplicity fluctuations.

From Eq.(1), one could obtain a fixed (mean) value of $dN^i/dy$ for each hadron species i. In the past simulations~\cite{Song:2013qma,Zhu:2016puf},  the Monte-Carlo event generator simultaneously generates many profiles from one switching hyper-surface with fixed number of multiplicity $N_{i}=w_{y}*dN^i/dy$  for each hadron species i (where $w_{y}$ is the width of the rapidity window),  which are then input into  {\tt UrQMD} for the  succeeding evolution of the hadronic matter. The multiplicity fluctuations of the final produced hadrons are mainly come from the initial state fluctuations and the fluctuations from the evolution, scatterings and decays of the hadronic matter.

In the this paper, we assume the emitted hadrons from the hydrodynamic switching hyper-surface contain additional thermal fluctuations that obey the Poisson distribution:
\begin{equation}
P_{i}(k)=\frac{\lambda_{i}^{k}e^{-\lambda_{i}}}{k!},
\end{equation}
Here, $\lambda_{i}=N_i$ which is the mean value (multiplicity) of the hadron species i. Note that the Copper-fryer freeze-out of hydrodynamics is belong to the framework of statistical hadronization with grand canonical ensemble. For heavier particles like protons, the equilibrium distribution $f_0(x,p)$ is very close to the Boltzmann distribution near $T_{sw}$, which leads to an approximately Poisson distribution after considering the related thermal fluctuations. We have also realized that other factors, such as the non-equilibrium distribution function $\delta f$, the Bose-Einstein distributions for light hadrons, etc. could break such Poisson distribution to some extend. As pointed out in~\cite{Shen:2014vra,ShenPhD}, an exact implementation of the realistic thermal fluctuations in {\tt iEBE-VISHNU} is non-trivial. Here, we take such distribution in Eq.(2) as a basic assumption, and then focus on investigating how the effects of volume fluctuations, hadronic evolution, resonance decays, etc., influence the multiplicity fluctuations of final produced hadrons. \\

For the investigation of multiplicity fluctuations, the numerical efficiency is one of the most important factors to be considered since millions of final particle profiles are needed for the analysis of higher moments of final produced hadrons.  In our simulations,  we first run {\tt MC-Glb} model
to generate millions of initial profiles, and then cut the ``centralities" according to the distributions of total initial entropy~\footnote{The ``centrality" cut here are different from the centrality definition of final multiplicities, since the Poisson distribution in Eq.(2) brings additional multiplicity fluctuations, which influences the centrality cut of final produced hadrons within an acceptance window.}.
Here, we divide the initial profiles into 20 ``centrality bins".  For each unit bin, we run one hydrodynamic simulation with a smoothed initial entropy density averaged from $N$ ($N=100\sim1000$) {\tt MC-Glb} profiles within that ``centrality",  which then follows with $N$ {\tt UrQMD} simulations for the succeeding hadronic evolution. In more details, one hydrodynamic simulation generate one switching hyper-surface, which gives the mean multiplicity  $\bar{N}_{i}$ for each thermal hadron species.  For the $\alpha$'s initial profile within a specific ``centrality",  we do not run
the time-consuming hydrodynamic simulation, but estimate the mean multiplicity of each hadron species by $s_{\alpha}\bar{N}_{i}/\bar{s}$ considering that the multiplicity is approximately proportional to the initial entropy for a hydrodynamic system  (where $\bar{s}$ is total entropy of the event-averaged initial profiles and $s_{\alpha}$ is total entropy of the $\alpha$-th event). Before the succeeding {\tt UrQMD} simulations for the $\alpha$'s event, we add additional Poisson fluctuations for each thermal hadron species i through Eq.(2) with the Poisson parameter set to
$\lambda_{\alpha;i} = s_{\alpha}\bar{N}_{i}/\bar{s}$.  With such method and a properly chosen acceptance cut for final produced particles, the effect of volume fluctuations~\cite{Jeon:2003gk,Skokov:2012ds,Xu:2016qzd,Xu:2016skm} have been included in our model simulations.

In the following calculations,  we set $\tau_0=0.6$ fm/c, $\eta/s=0.08$, and neglect the bulk viscosity and heat conductivity~\cite{Song:2013gia,Song:2017wtw}.
We use the multiplicities, particle ratios, and the mean values of (anti-)protons, (negative) positive charges~\cite{Abelev:2008ab,Adamczyk:2013dal,Adamczyk:2014fia}
to fix the freeze-out parameters in Eq.(1), including the chemical freeze-out
temperature ($T_{ch}$), baryon chemical potential ($\mu_B$), strangeness chemical
potential ($\mu_S$), charge chemical potential ($\mu_Q$), and strangeness saturation
factor ($\gamma_s$). These fine tuned parameters are listed in Table~\ref{parameters}.

\begin{figure*}[!hbt]
	\begin{centering}
		\includegraphics[scale=0.445]{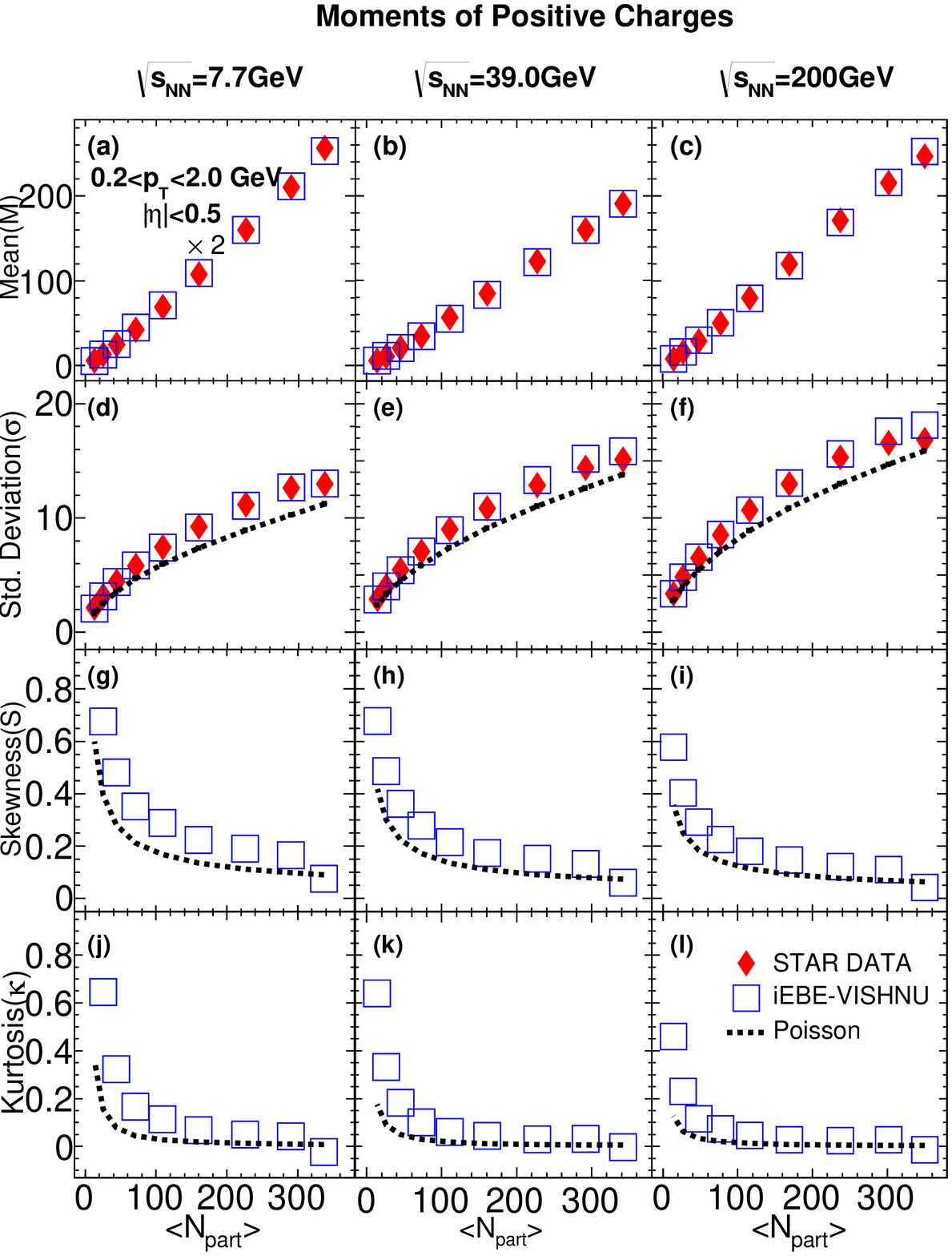}
		\includegraphics[scale=0.445]{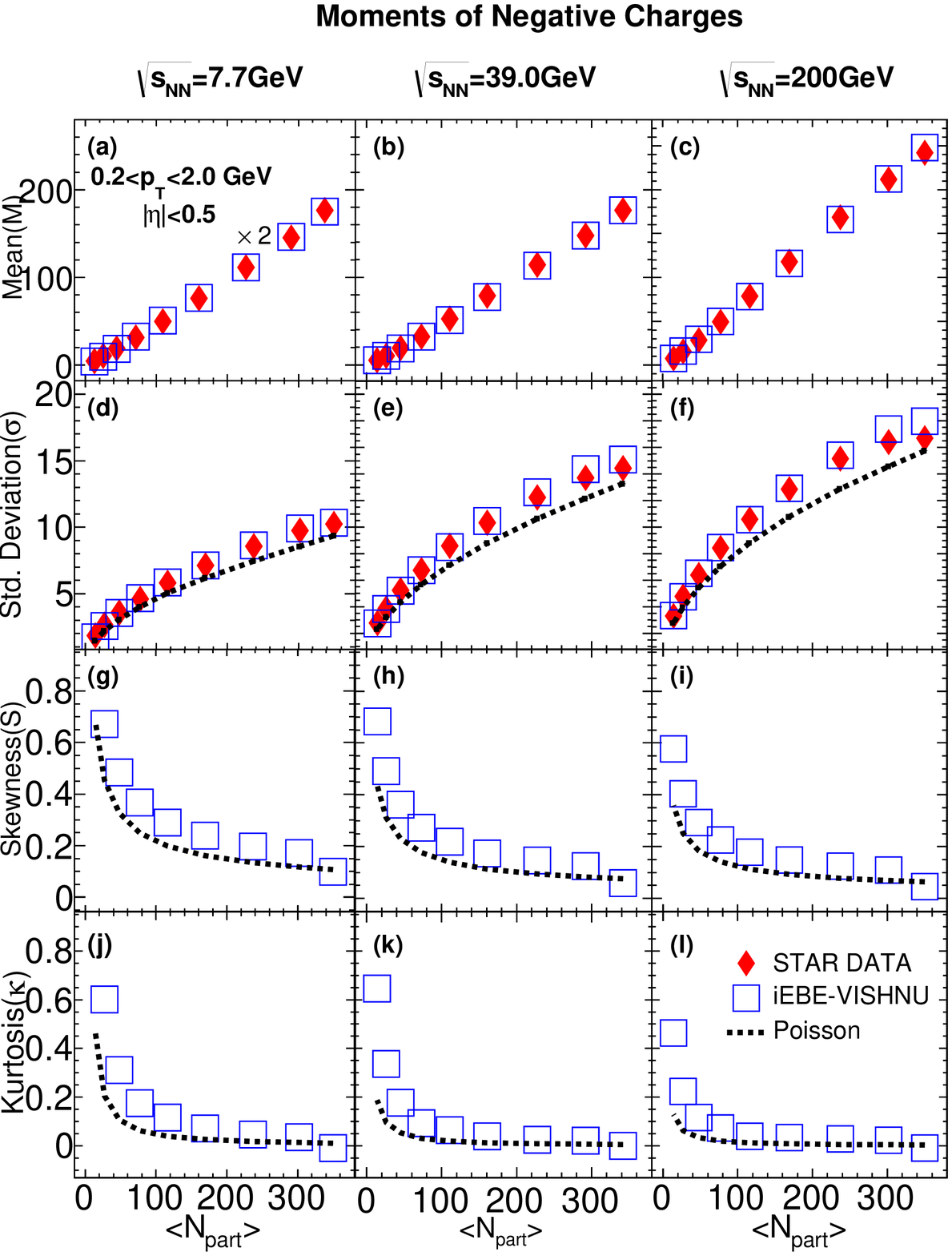}
	\end{centering}
	\vspace{-3mm}
	\caption{(Color online) Moments of positive and negative charges as a function of participant
number $\langle N_{\mathrm{part}} \rangle$ in Au+Au collisions at $\sqrt{S_{\text{NN}}}=7.7,39.0~\mathrm{and} ~200$ GeV. The theoretical results are calculated from {\tt iEBE-VISHNU}.  The data of mean value $M$ and standard variance $\sigma$ are from the STAR paper~\cite{Adamczyk:2014fia}. The dashed black lines are the Poisson baselines.
		\label{fig1:Posi_Nega_moments}}
\end{figure*}

\section{Observables}

%After the event-by-event simulations described above, we analyze the {\tt iEBE-VISHNU} results
%as did in experiments ~\cite{Adamczyk:2013dal,Adamczyk:2014fia}.
To evaluate the multiplicity fluctuations,
one calculates the cumulants of the multiplicity distributions of final produced particles:
\begin{align}\label{Cumulants}
	c_{1} &= \langle N\rangle\equiv M,  \tag{3-a}\\
	c_{2} &= \langle (\Delta N)^{2}\rangle \equiv \sigma^{2}, \tag{3-b}\\
	c_{3} &= \langle (\Delta N)^{3}\rangle \equiv S\sigma^{3}, \tag{3-c}\\
	c_{4} &= \langle (\Delta N)^{4}\rangle - 3 c_{2}^{2} \equiv \kappa\sigma^{4}, \tag{3-d}
\end{align}
where $\Delta N = N-\langle N\rangle$, $N$ is
the multiplicity of the particle of interest, and
$\langle...\rangle$ denotes the event average. Here, $M$, $\sigma$,
$S$ and $\kappa$ are the mean value, standard variance, skewness and kurtosis
of the probability distribution. In order to partially remove the
volume effects, one calculates the moment products $\sigma^2/M$, $S\sigma$ and $\kappa\sigma^2$, which can be expressed as the cumulant ratios:
\begin{eqnarray}\label{momentsRatio}
 \sigma^2/M = \frac{C_2}{C_1},~~~ S\sigma = \frac{C_3}{C_2}, ~~~\kappa\sigma^{2}= \frac{C_4}{C_2},
\end{eqnarray}

In experiments, the multiplicity fluctuations of (net) charges and (net) protons are measured within certain centrality bin and
acceptance window. For positive and negative charges, one first cuts the centrality bin by the total number of all charged hadrons within the pseudo-rapidity window  $0.5 < |\eta| < 1.0$, and then measures the event-by-event multiplicity distributions of  positive and negative charges within the acceptance cut $|\eta| < 0.5$ and $0.2 < p_T < 2.0$ GeV for each centrality bin.
For the case of protons and anti-protons,  the centrality bin is cut by the total number of pions and kaons within the pseudo-rapidity window  $ |\eta| < 1.0$. For each centrality, the multiplicity fluctuations of protons and anti-protons are analyzed within the mid-rapidity $|y|<0.5$ and transverse momentum $0.4 < p_T < 0.8$ GeV.
Follow the experiments~\cite{Adamczyk:2013dal,Adamczyk:2014fia}, we use the same centrality definition, acceptance cut, as well as the following centrality bin width corrections in our model calculations.

In general, the multiplicity fluctuations are not directly measured within a wide centrality bin (e.g. 0-5\%, 10-20\%, etc.), which associates with the wide centrality bin effects that can distort the imprinted fluctuations~\cite{Luo:2011ts}. To reduce such effects, one divides a wide centrality bin into many fine bins, and then calculates the total moments within that wide centrality bin from the moments of each fine bin with some weight:
\begin{equation}
X=\frac{\sum_i n_i X_i}{\sum_i n_i },
\label{eq:bin_with}
\end{equation}
where $X$ represents the total moment within a wide centrality, $X_i$ is the moment of the fine centrality bin $i$.  $n_i$ is the number of events in the $i^{th}$ bin, and $\sum_i n_i$ is the total number of events in the wide centrality bin. Following~\cite{Luo:2014rea,Luo:2011tp}, we choose each reference multiplicity to define a fine centrality bin~\cite{Adamczyk:2013dal}, and then implement the Delta theorem to calculate the statistical errors for the moments and moment products.\\[0.20in]
\begin{figure}[tbh]
	\begin{centering}
		\includegraphics[scale=0.445]{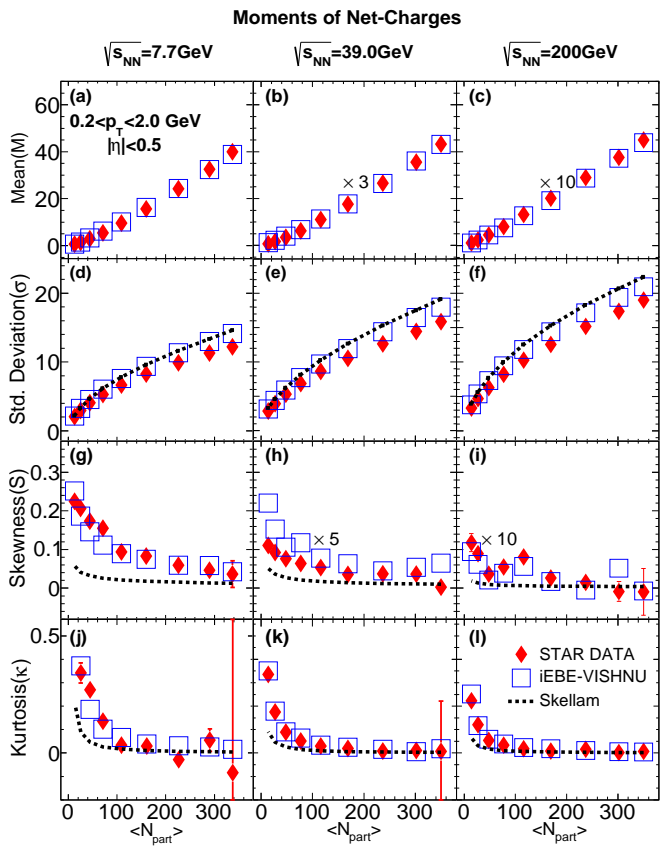}	
	\end{centering}
	\vspace{-3mm}
	\caption{(Color online) Similar to Fig.~1, but for the moments of net-charges, calculated from
    {\tt iEBE-VISHNU} and measured by STAR~\cite{Adamczyk:2014fia}.  The dashed black lines are the Skellam baselines.
    }
\end{figure}

\begin{figure}[htb]
    \vspace{3mm}
	\begin{centering}
    \includegraphics[scale=0.45]{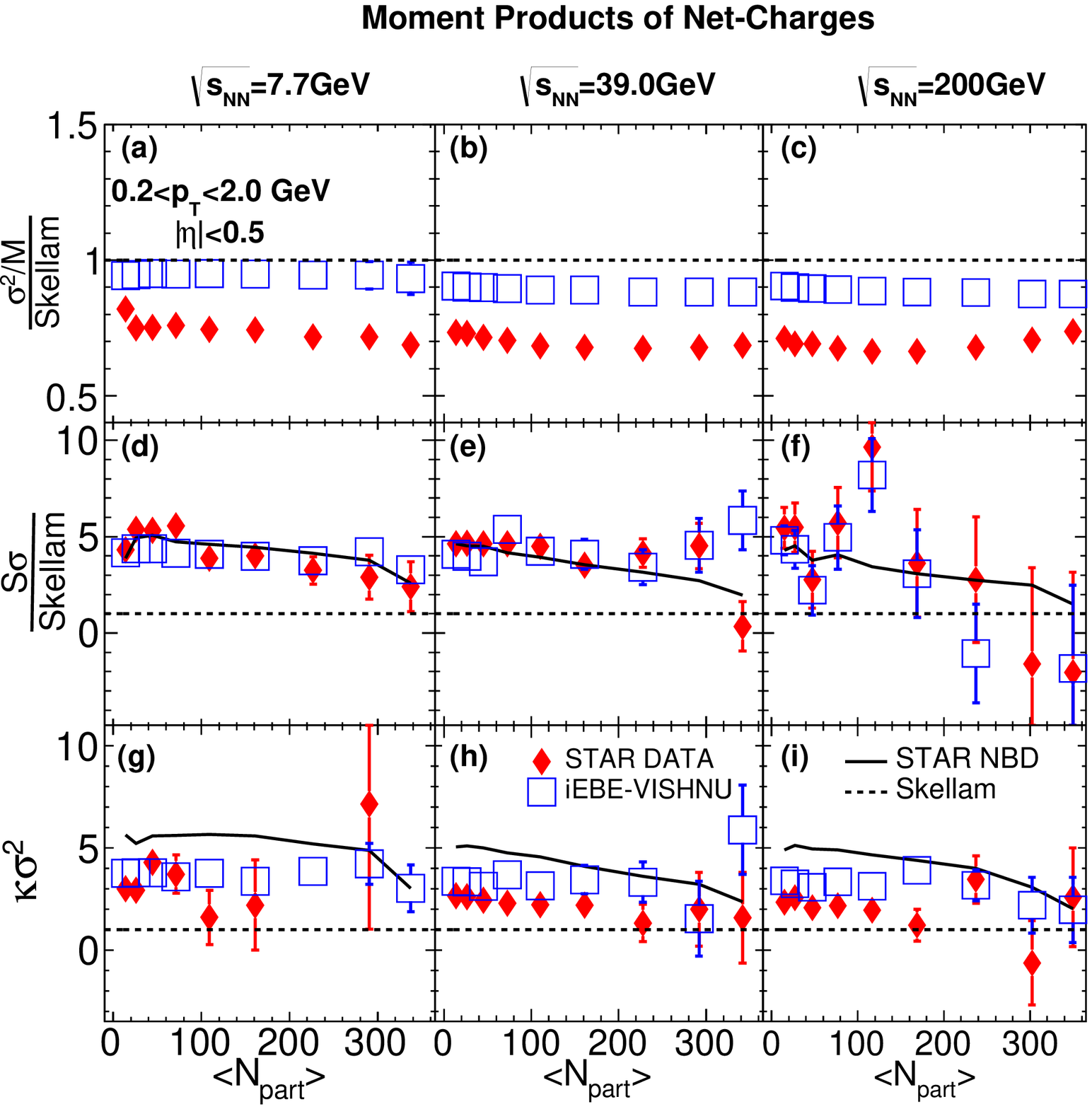}	
	\end{centering}
	\vspace{-3mm}
	\caption{(Color online) Centrality dependent moment products $\frac{\sigma^{2}/M}{\mathrm{Skellam}}$,
    $\frac{S\sigma}{\mathrm{Skellam}}$ and $\kappa\sigma^2$ of net-charges in Au+Au collisions at $\sqrt{S_{\text{NN}}}$ =
      7.7, 39.0 and 200 GeV, calculated from {\tt iEBE-VISHNU} and measured by STAR~\cite{Adamczyk:2014fia,Adamczyk:2013dal}.
      \label{fig3:NetProtonChargeMoments}}
\end{figure}
\begin{figure*}[htbp]
	\begin{centering}
		\includegraphics[scale=0.4450]{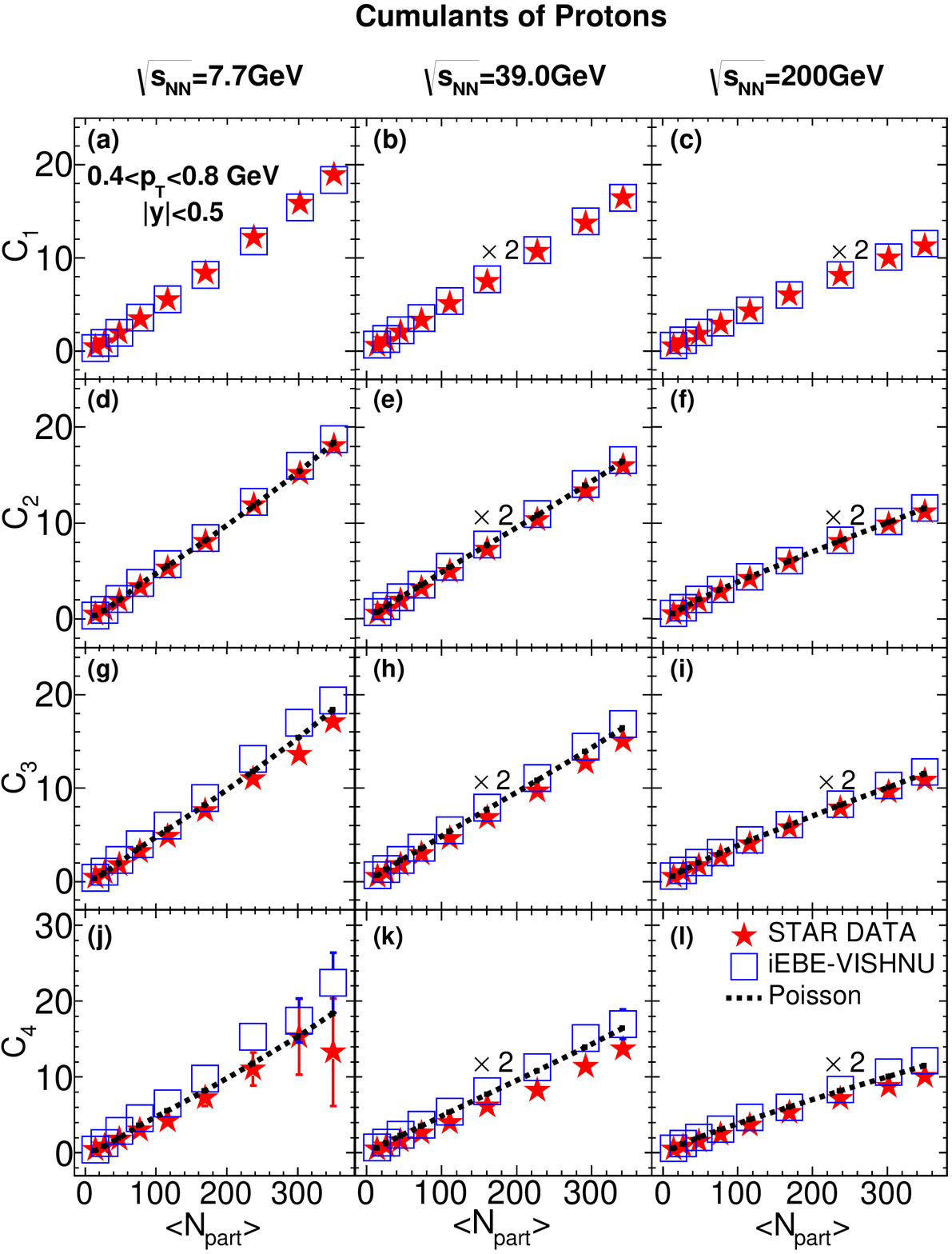}
		\includegraphics[scale=0.4450]{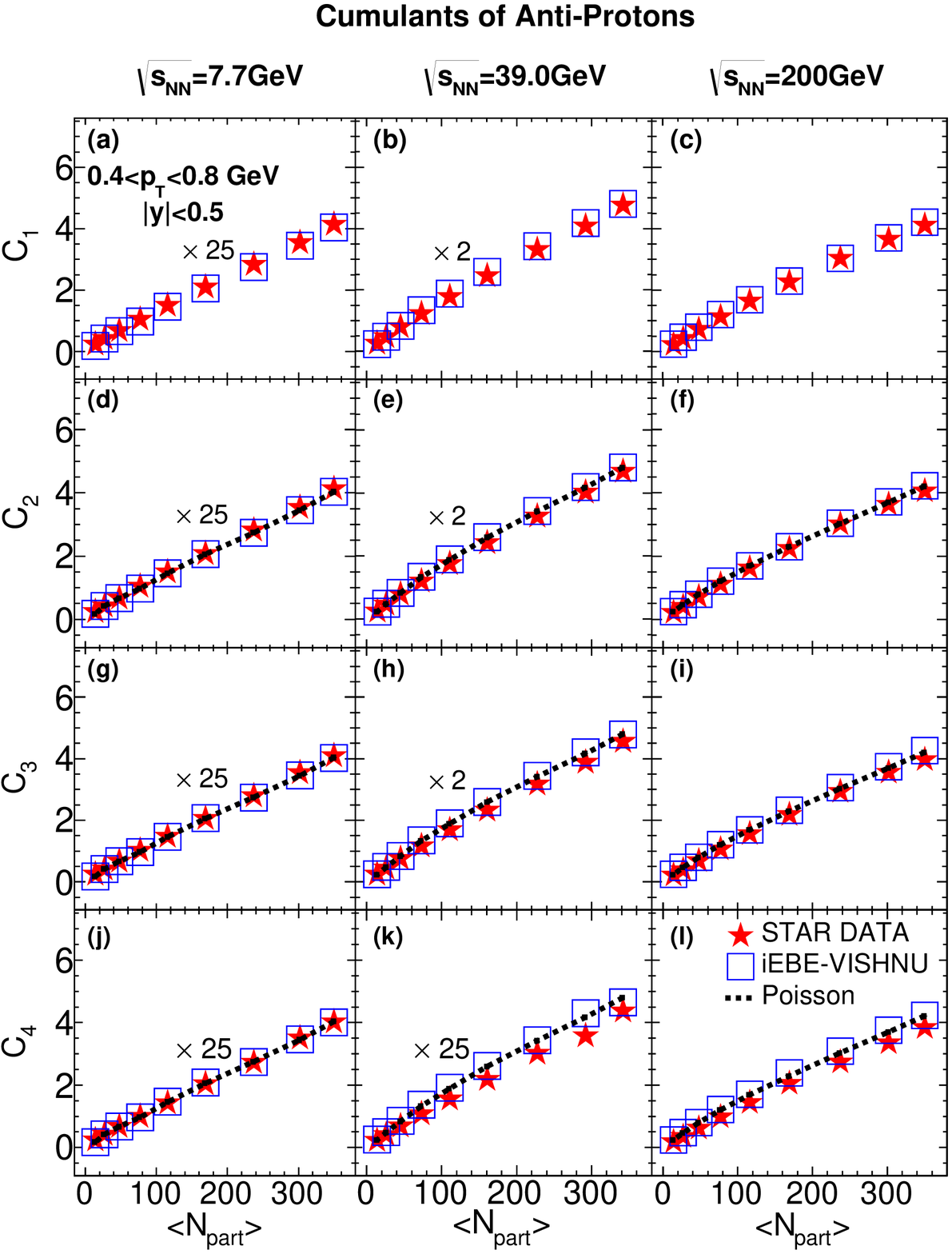}
	\end{centering}
	\vspace{-2mm}
	\caption{(Color online) Cumulants of protons and
    anti-protons as a function of
    participant number $\langle N_{\mathrm{part}} \rangle$  in Au+Au collisions at
    $\sqrt{S_{\text{NN}}}=7.7, 39~ \mathrm{and} ~200$ GeV, calculated from {\tt iEBE-VISHNU} and measured by
    STAR~\cite{Adamczyk:2013dal}. The dashed black lines are the Poisson baselines.
    \label{fig4:Pro_Anti_Proton_Cumulants}}
\end{figure*}

\begin{figure}[htbp]
	\begin{centering}
		\includegraphics[scale=0.45]{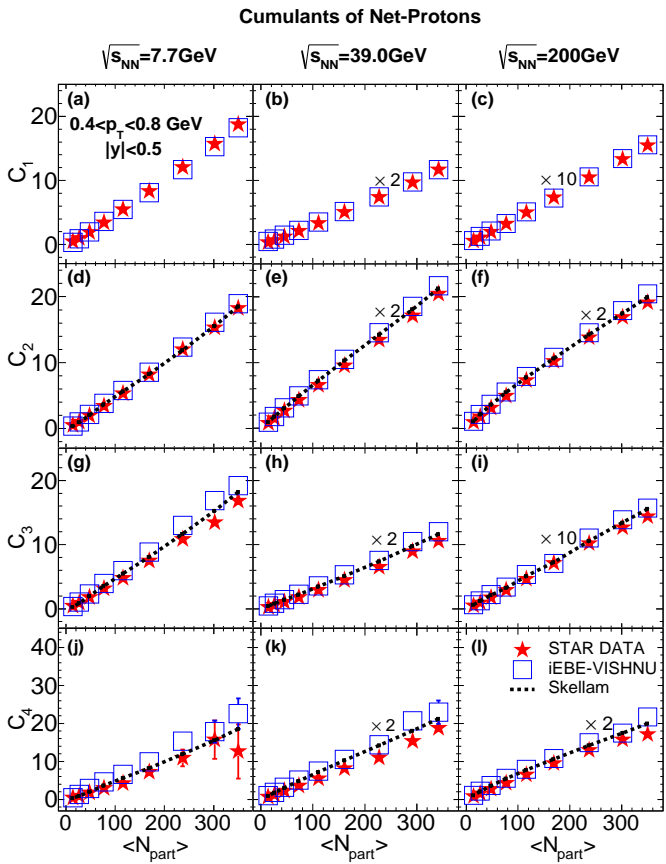}
	\end{centering}
	\vspace{-2mm}
	\caption{(Color online) Similar to Fig.~4, but for the cumulants
       of net protons.  The dashed black lines are the Skellam baselines.
     \label{fig5:Net_Proton_Cumulants}}
\end{figure}

\begin{figure}[htb]
   \vspace{3mm}
	\begin{centering}
		\includegraphics[scale=0.445]{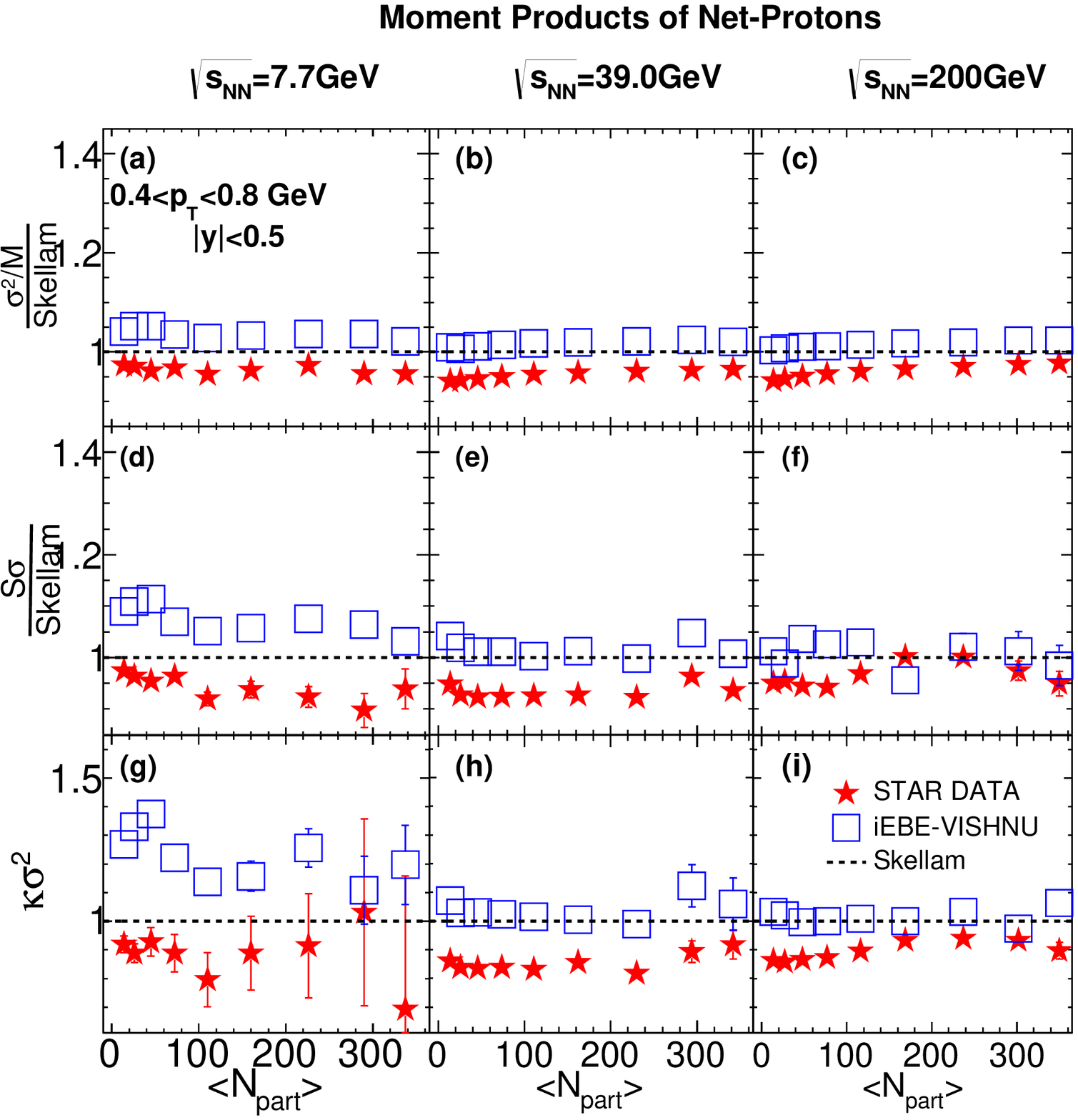}
	\end{centering}
	\vspace{-2mm}
	\caption{(Color online) Centrality dependent moment products $\frac{\sigma^{2}/M}{\mathrm{Skellam}}$,
    $\frac{S\sigma}{\mathrm{Skellam}}$ and $\kappa\sigma^2$  of net-protons in Au+Au collisions at $\sqrt{S_{\text{NN}}}$=7.7,
      39.0 and 200 GeV, calculated from {\tt iEBE-VISHNU} and measured by
    STAR\cite{Adamczyk:2014fia,Adamczyk:2013dal}. The dashed black lines are the Skellam baselines.
      \label{fig6:NetProtonChargeMoments}}
\end{figure}

\section{Results and discussions} \label{results}
\subsection[Ａ]{Comparisons with the STAR data  }
Fig.~1 shows the centrality dependent moments (mean value $M$, standard deviation $\sigma$, skewness $S$ and
kurtosis $\kappa$) of positive and negative charges in Au + Au collisions at $\sqrt{s_{\mathrm{NN}}}= 7.7$,
$39$ and  $200$ GeV.  In our model calculations, we first tune the related parameters in {\tt iEBE-VISHNU} to fit the mean values of positive and negative charges, and then predict other  moments at these selected collision energies (please refer to Sec. II for details).  For the centrality-dependent standard variance $\sigma$, {\tt iEBE-VISHNU} nicely describe the data of both positive and negative charges, except for the most-central collisions. Meanwhile, our model calculations show certain deviations from the Poisson baselines for various moments, including the standard deviation $\sigma$, skewness $S$ and kurtosis $\kappa$.

In {\tt iEBE-VISHNU} calculations, there are many factors that could influence the multiplicity fluctuations of final produced hadrons, which include the initial state fluctuations, the poisson fluctuations for the emitted hadrons on the hydrodynamic switching surface, the hadronic scatterings and resonance decays in {\tt UrQMD},  as well as the centrality and acceptance cuts for the particle of interest.  In the following Sec IV B, we will show that the effects from the hadronic evolution and decays are pretty small or even negligible for the multiplicity fluctuations of (net) charges and (net) protons. For a certain centrality bin, the combined effects of initial entropy fluctuations and the Poison fluctuations on the switching surface are similar to the volume fluctuations/corrections as investigated in the early paper~\cite{Skokov:2012ds,Xu:2016qzd}, which are the dominant factors to influence the multiplicity fluctuations and make them deviate from the Poison Baselines (please refer to Sec. IV B for details).

We also noticed that, in the most central collisions, the standard deviation $\sigma$ of {\tt iEBE-VISHNU} is about 10\% higher than the experimental data.  In~\cite{Skokov:2012ds,Xu:2016qzd}, it was found that
the effects of volume fluctuations are largely suppressed in the
most-central collisions. Correspondingly,  the multiplicity fluctuations there are more sensitive to other factors, such as the initial state fluctuations, resonance decays, and etc. In Fig.~1, the slightly over-predictions of the data at 0-5\% centrality indicates that the used {\tt MC-Glb} initial conditions may not fully capture the fluctuation patterns as imprinted in nature.
More sophisticated model calculations, especially for the most central collisions, are still needed which we would like to leave it to the future study.

Fig.~2 shows the centrality dependent moments
of net-charges in Au+Au collisions at $\sqrt{s_{NN}} =7.7$,
$39$ and $200$ GeV. Although {\tt iEBE-VISHNU} has archived an
overall fit of the standard variance $\sigma$ for both positive and negative charges,
it fails to nicely describe the corresponding $\sigma$ of net-charges, which shows certain
deviations between model and data.  For the skewness $S$
and kurtosis $\kappa$ of net-charges,  {\tt{iEBE-VISHNU}} quantitatively describes the data within the
statistical errors. In contrast, the Skellam baselines (which come from the subtraction of two
independent Poisson distributions) show certain deviations from the experimental data,
especially at lower collision energies. In~\cite{Xu:2016skm}, it was found that
the volume fluctuations/corrections are the dominant factors to influence the skewness $S$ and
kurtosis $\kappa$ of net charges, but are negligible for the corresponding standard variance $\sigma$. This leads to the the different descriptions of $\sigma$, $S$ and $\kappa$
in our model calculations, which will be further discussed in the following Sec.~IV B.

Besides the effects of volume fluctuations, the correlation between positive
and negative charges is  another important factor to influence the fluctuations of net charges, which may
even play a dominant role to affect the standard variance $\sigma$ of net charges. In  {\tt{iEBE-VISHNU}} model,
such correlations mainly come from the resonance decays during the late hadronic evolution. However, some
additional correlations, e.g. the correlations from the charge conservation laws, are still missing,
which may largely influence the standard variance $\sigma$ of net charges and should be
investigated in the near future.

Fig.~3 shows the centrality dependent moment products $\frac{\sigma^{2}/M}{\mathrm{Skellam}}$, $\frac{S\sigma}{\mathrm{Skellam}}$ and $\kappa\sigma^{2}$ of net-charges in Au+Au collision at $\sqrt{s_{NN}}=7.7$, $39$ and $200$ GeV.  In general, {\tt iEBE-VISHNU} roughly describes these experimental data. The slight deviations mainly come from the over-predictions of the standard variance $\sigma$ of net-charges (please refer to Fig.~2 and the related discussions).
Note that part of the correlations between positive
and negative charges, e.g. from resonance decays and hadronic scatterings, has been included in our calculations, which makes the {\tt iEBE-VISHNU} results are more close to the STAR data than the negative
binomial baselines~\footnote{{As shown in Ref.~\cite{Xu:2016qzd}, negative binomial baselines can be obtained from the Poisson distribution after considering the effects of volume fluctuations with some approximations.}}.\\

With the same parameter sets, we calculate the cumulants ($C_1 - C_4$) of protons, anti-protons and net-protons in Au+Au collisions at $\sqrt{s_{NN}}=7.7$, $39$ and $200$ GeV. Fig.~4 and fig.~5 show that our {\tt{iEBE-VISHNU}} results are pretty close to the experimental data, which all monotonically increase with the participant number $\langle N_{part}\rangle$.  We also notice that the difference between our model calculations and the Poisson baselines are pretty small, which indicates that various effects included in our model calculations, i.e. volume fluctuations,  hadronic scatterings  and resonance decays, do not significatively influence the cumulants of protons, anti-protons and net protons. For more detailed discussions,  please also refer to Sec IV B.

For a closer look,  figure~6 plots the moment products
$\frac{\sigma^{2}/M}{\mathrm{Skellam}}$, $\frac{S\sigma}{\mathrm{Skellam}}$ and $\kappa\sigma^{2}$  of net protons,  which presents certain deviations between the data and
our model calculations.  Compared with the Skellam baselines that are obtained from the two independent Poisson distributions of protons and anti-protons,  most of the {\tt iEBE-VISHNU} results and the experimental data
are respectively below and above the baselines.  Fig.~6 also shows that the gap between model calculations and Skellam baselines increase with the decrease of collision energies. As discussed in Ref.~\cite{Xu:2016skm} and Sec.IV B, the related effects of volume fluctuations are closely related to the value of  $(M_+-M_-)/(k+1)$ (where $M_+-M_-$ is the mean value of net protons and $k$ is the reference multiplicity), which increases with the decrease of collision energy and leads to the increasing gap between our model calculations and the Skellam baselines.

In many past research~\cite{Jiang:2015hri,Jiang:2017mji}, the cumulant ratios of net protons are expected as sensitive observables to probe the non-gaussian fluctuations of the QCD critical point and the first order phase transitions. Note that our model calculations only include the various effects of non-critical fluctuations. The failure of describing the cumulant ratios of net protons at lower collision energies indicates that other possible effects, such as baryon conservation laws, critical fluctuations and spinodial instabilities of the first order phase transitions, may largely influence the multiplicity fluctuations of net protons there, which are worthwhile to be further studied in the near future.

\begin{figure*}[htb]
	\begin{centering}
		\includegraphics[scale=0.43]{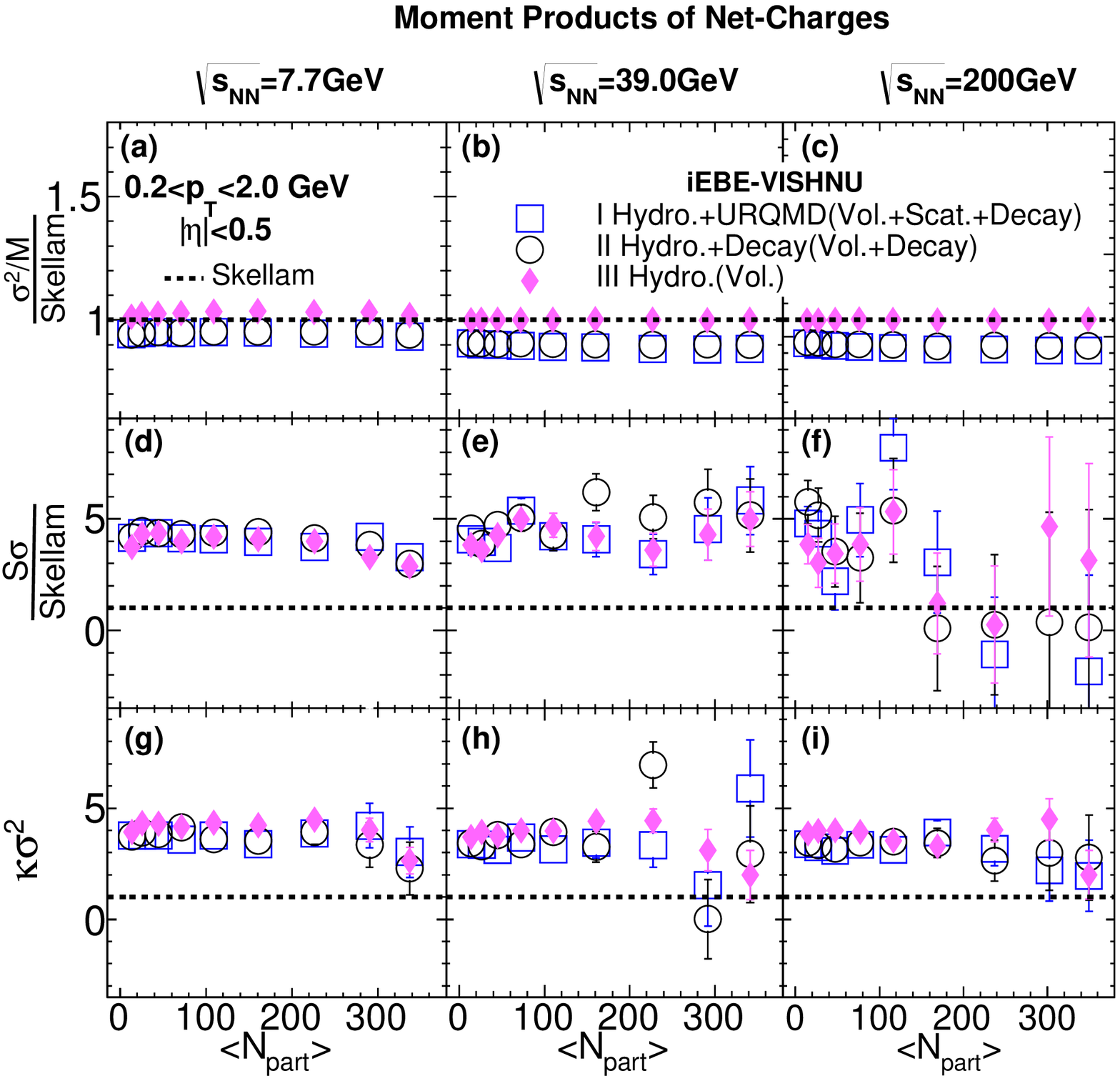}
		\includegraphics[scale=0.43]{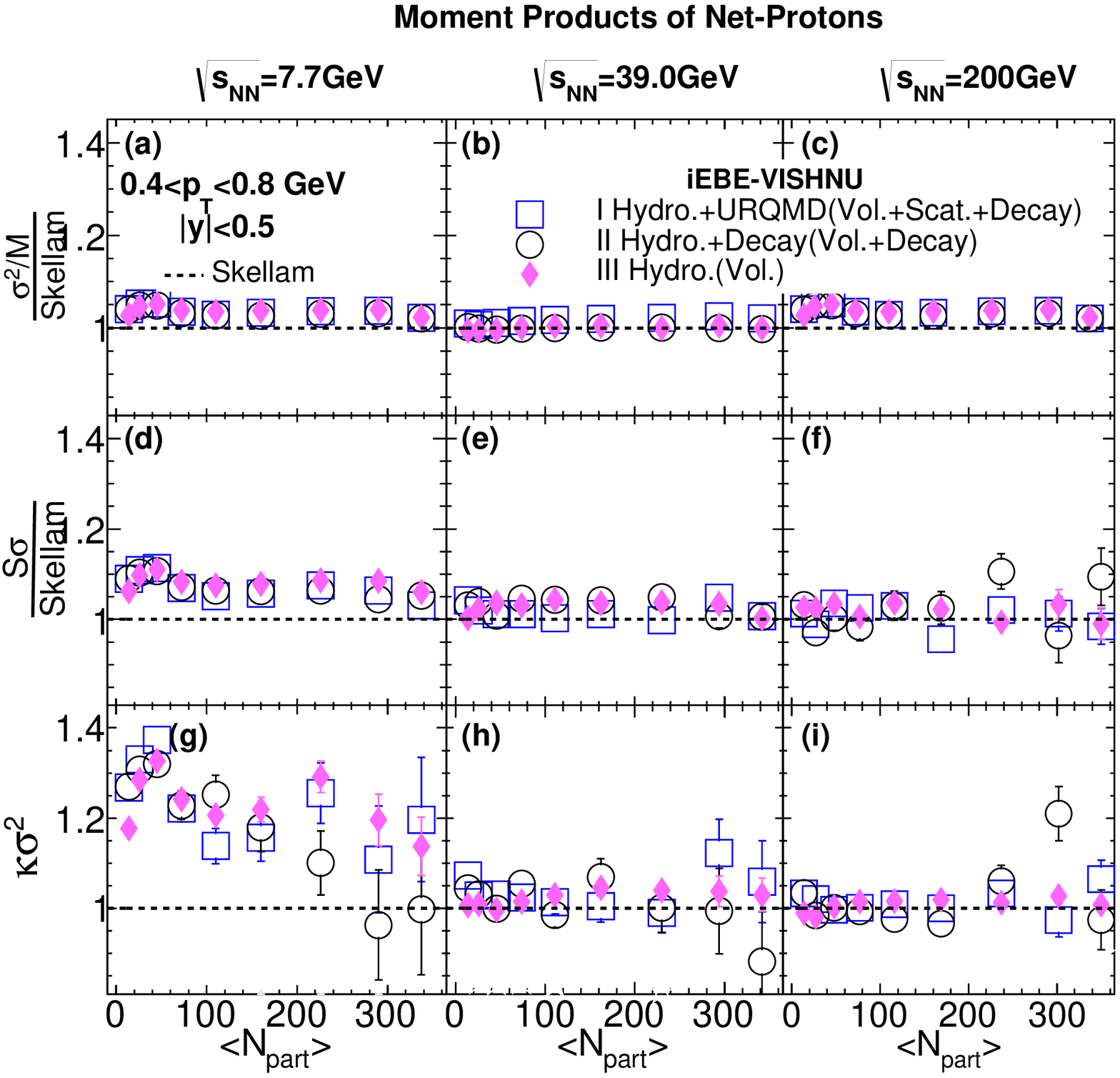}
	\end{centering}
	\vspace{-2mm}
	\caption{(Color online) Moment products $\frac{\sigma^{2}/M}{\mathrm{Skellam}}$, $\frac{S\sigma}{\mathrm{Skellam}}$ and $\kappa\sigma^2$ of net charges and net protons in Au+Au collisions at $\sqrt{S_{\text{NN}}}$=7.7, 39 and 200 GeV, calculated from {\tt iEBE-VISHNU} with three different simulation modes.
\label{fig8_NetProtonChargeDynamic}}
\end{figure*}

\subsection[B]{The effects of volume fluctuations, resonance decays and hadronic evolution}
As we have mentioned, various effects such as volume fluctuations, hadronic scatterings, resonance decays, etc., could influence the multiplicity fluctuations of (net) charges and (net) protons to some extend. To further explore these effects, we perform the model simulations with three different cases: \textbf{(a)} full {\tt{iEBE-VISHNU}} simulations as did in Sec.IV A, where the hydrodynamic expansion is followed by a full {\tt UrQMD} hadronic evolution with both hadronic scatterings and decays; \textbf{(b)} modified {\tt{iEBE-VISHNU}} simulations with hydrodynamics followed by resonance decays, but without the hadronic scatterings and evolution of {\tt UrQMD}; \textbf{(c)} pure hydrodynamic simulations with thermal hadrons directly emitted from the freeze-out hyper-surface with the imprinted Poisson fluctuations described by Eq.~(2).

Fig.~7 presents the centrality dependent moment products $\frac{\sigma^{2}/M}{\mathrm{Skellam}}$, $\frac{S\sigma}{\mathrm{Skellam}}$ and $\kappa\sigma^{2}$ of net-charges and net-protons in Au+Au collisions at $\sqrt{S_{\text{NN}}}$=7.7, 39 and 200 GeV, obtained from {\tt {iEBE-VISHNU}} simulations with these three above modes.   For the simulations with case \textbf{(c)}, we focus on the effects of the volume fluctuations.  More specifically, for a single hydrodynamic simulation, thermal hadrons directly emitted from the freeze-out surface satisfy the Poisson fluctuations according to Eq.(2), which also generates the corresponding Skellam baselines in Fig.~7.  In the event-by-event simulations, a specific reference multiplicity $k$ (within a centrality cut window e.g. $0.5 < |\eta| < 1.0$) can be generated from many different hydrodynamic simulations with the
initial state fluctuations and poisson fluctuations.  The related multiplicity fluctuations (within an acceptance cut window e.g. $0< |\eta| < 0.5$) for a given reference multiplicity bin thus surfer similar volume fluctuation effects as investigated in early paper~\cite{Xu:2016qzd, Xu:2016skm}.

Fig.~7 shows that, the volume fluctuations/corrections for $\sigma^{2}/M$ of net charges are pretty small or even negligible, but very large for $S\sigma$ and $\kappa \sigma^2$ which make them obviously deviate from the Skellam baselines. For net protons, the volume fluctuations/corrections for $\sigma^{2}/M$, $S\sigma$ and $\kappa \sigma^2$ are all pretty small, but also gradually increase with the decrease of collision energy. In~\cite{Xu:2016skm}, it was found that the volume fluctuations/corrections for  the standard variance $\sigma$ of net charges (protons) are approximately proportional to $(M_+-M_-)/(k+1)$, where $M_{+}$ and $M_-$ are the mean multiplicity of positive and negative charges (protons) and $k$ is the reference multiplicity for the centrality cut.  Note that, for both net charges and net protons $(M_+-M_-) \ll (k+1)$. This largely suppresses the related volume fluctuations of $\sigma$, which is also directly demonstrated in our model calculations of Fig.~7.   For the skewness $S$ and kurtosis $\kappa$, the volume correction is not only dependent on $(M_+-M_-)/(k+1)$, but also depend on $(M_++M_-)/(k+1)$~\cite{Xu:2016skm}. For net charges, $(M_++M_-)/(k+1)\sim 1$, we thus observe large volume fluctuations for both $S\sigma$ and $\kappa \sigma^2$ in Fig.~7 (left). For the case of net protons, $(M_++M_-)/(k+1)$ are still pretty small, but gradually increase with the decrease of collision energy.  Correspondingly, we find the {\tt iEBE-VISHNU} results with only volume fluctuations are still pretty close to the Skellam baselines for both $S\sigma$ and $\kappa \sigma^2$. Meanwhile, the increased mean values of protons and anti-protons also leads to slightly larger deviations from the Skellam baselines for $S\sigma$ and $\kappa \sigma^2$ at lower collision energies.

Fig.~7 also compares the model simulations with case \textbf{(a)} \textbf{(b)} and \textbf{(c)}.  For $S\sigma$ and $\kappa\sigma^{2}$ of net-charges,  the results from the three comparison runs almost overlap within error bars, which also obviously deviate from the Skellam baselines.  This indicates that the volume fluctuations are the dominant factors to influence these two moment products of net-charges.  For $\sigma^2/M$, the volume fluctu are largely suppressed as discussed above. Fig.~7 shows that resonance decays become the dominant roles to influence $\sigma^2/M$, while the effects from the hadronic evolution and scatterings are pretty small.  For the moment products of net protons in Fig.~7 (right), both the resonance decays and hadronic evolution do not significantly influence the values of $\sigma^2/M$, $S\sigma$ and $\kappa\sigma^{2}$. In general, the effects of volume fluctuations are also pretty small (except for 7.7 GeV), which indicates that the multiplicity fluctuations of (net) protons keep the main features of the Poison fluctuations imprinted in our model calculations.

\begin{figure*}[htb]
	\begin{centering}
		\includegraphics[scale=0.42]{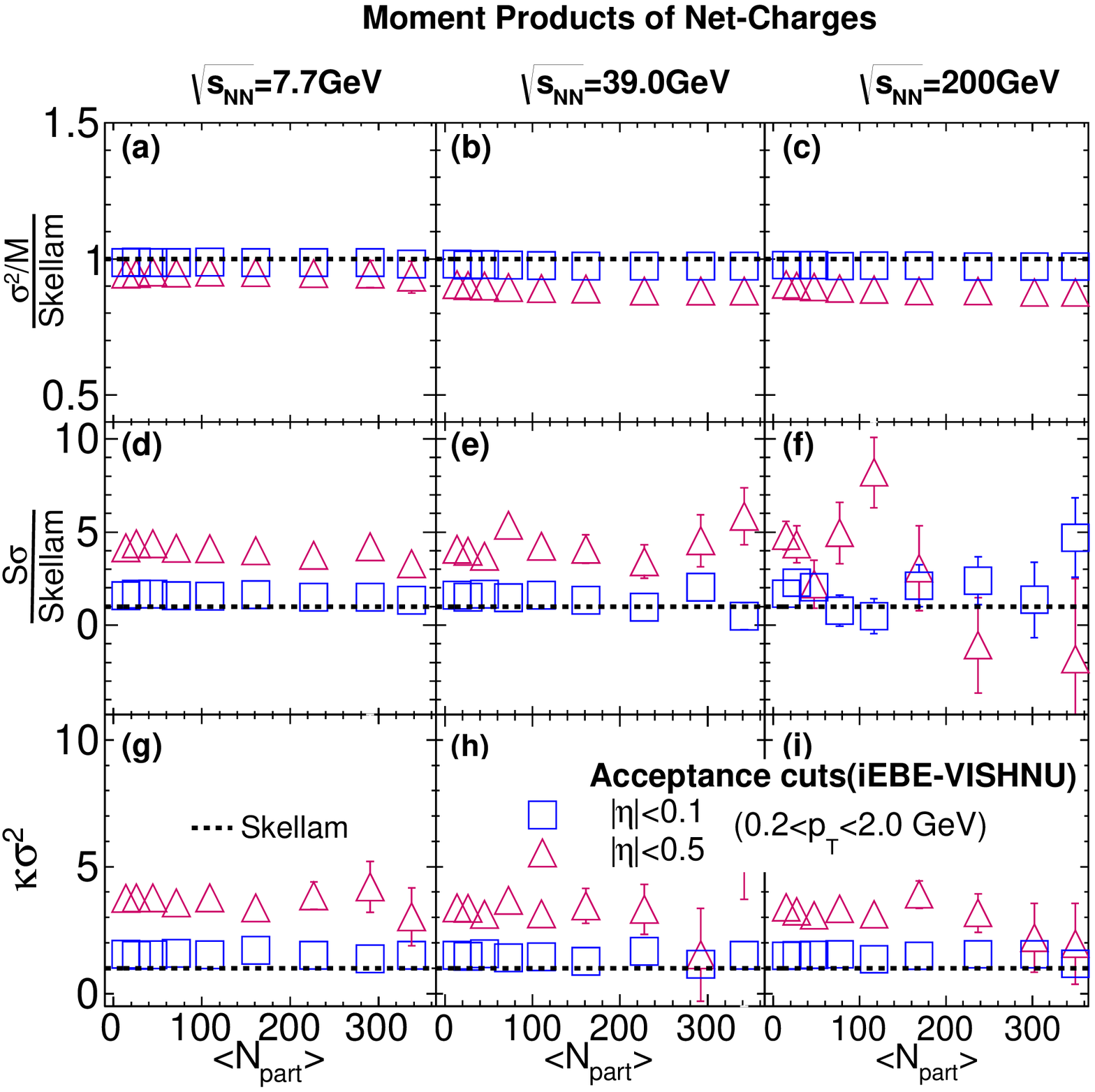}
		\includegraphics[scale=0.42]{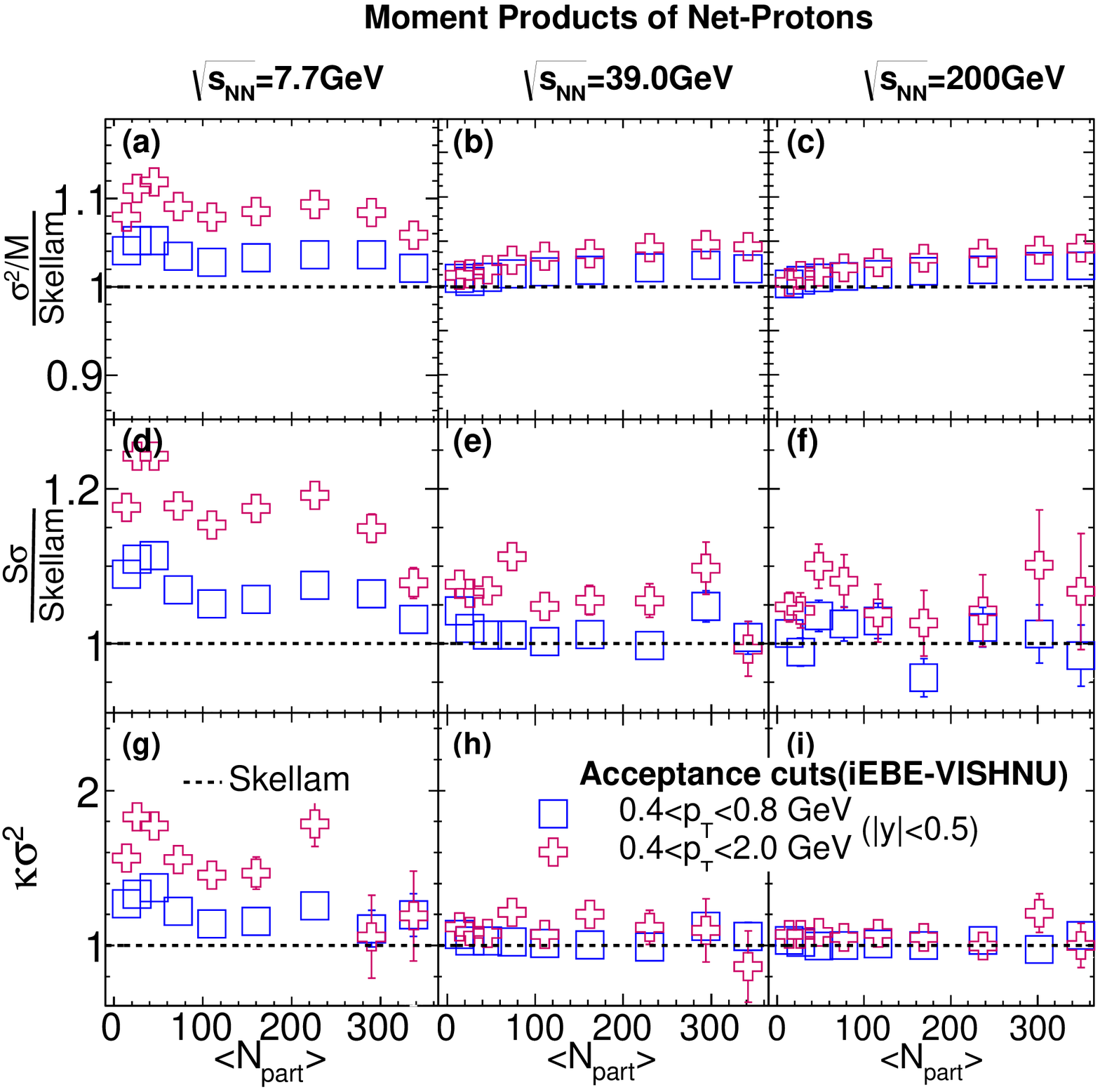}
	\end{centering}
	\vspace{-2mm}
	\caption{(Color online) Acceptance cut dependence of the moment products $\frac{\sigma^{2}/M}{\mathrm{Skellam}}$, $\frac{S\sigma}{\mathrm{Skellam}}$ and $\kappa\sigma^2$  in Au+Au collisions at $\sqrt{S_{\text{NN}}}$=7.7, 39 and 200 GeV, calculated from {\tt iEBE-VISHNU} with the acceptance cut set to $|\eta|<0.1$ and $|\eta|<0.5$ (with $0.4<p_T<2$ GeV)  for net charges and $0.4<p_T<0.8$ GeV and $0.4<p_T<2$ GeV (with $|y|<0.5$)  for net protons.
 \label{fig9_Acceptance}}
\end{figure*}

\begin{figure*}[htb]
	\begin{centering}
		\includegraphics[scale=0.42]{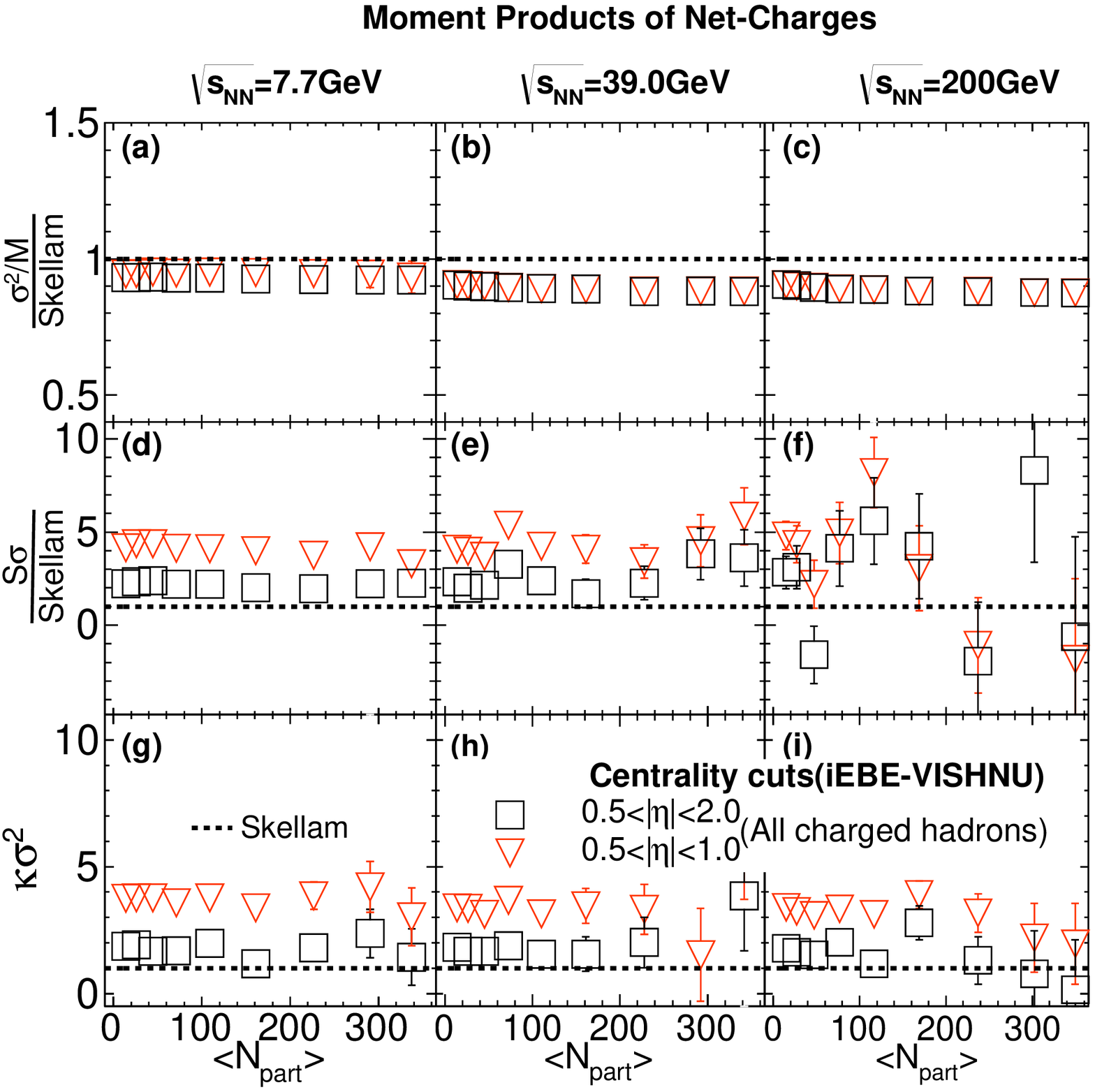}
		\includegraphics[scale=0.42]{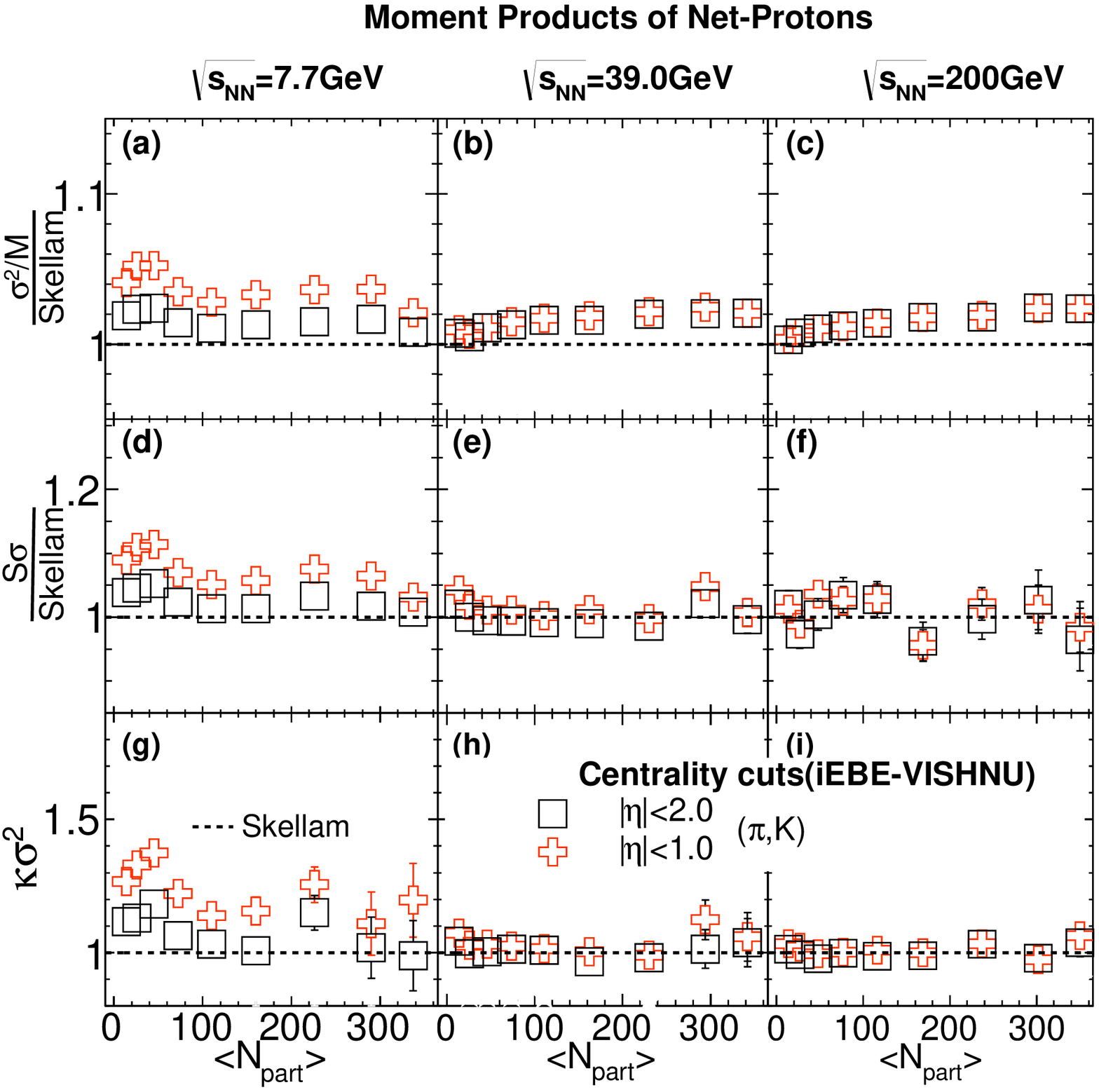}
	\end{centering}
	\vspace{-2mm}
	\caption{(Color online) Centrality cut dependence of the moment products $\frac{\sigma^{2}/M}{\mathrm{Skellam}}$, $\frac{S\sigma}{\mathrm{Skellam}}$ and $\kappa\sigma^2$  in Au+Au collisions at $\sqrt{S_{\text{NN}}}$=7.7, 39 and 200 GeV, calculated from {\tt iEBE-VISHNU} with the centrality defined by total charges  within $0.5<|\eta|<1.0$ or $0.5<|\eta|<2.0$ (left panels, for net charges) and by total pions and kaons within $|\eta|<1.0$ or $|\eta|<2.0$ (right panels, for net protons).
 \label{fig9_Acceptance}}
\end{figure*}

\subsection[C]{The dependence on centrality and acceptance cuts}

In Fig.~8, we study the acceptance  dependence of the moments products of
net-charges and net-protons, using the {\tt iEBE-VISHNU} simulations with different pseudo-rapidity and transverse momentum cut.  We find broader acceptance windows lead to larger deviations from the Skellam baselines.  As discussed above, the volume fluctuation in our model calculations is the main factor to influence the multiplicity fluctuations of net charges and net protons, especially for moments products of Skewness and Kurtosis. For a broader acceptance window,  the mean values of net charges and net protons $(M_+-M_-)$ increases. This leads to larger values of $(M_+-M_-)/(k+1)$ for a fixed reference multiplicity bin $k$, which enhances the corresponding volume fluctuations for $\sigma^{2}/M$, $S\sigma$ and $\kappa\sigma^{2}$, making them deviate from the Skellam baselines

We have also noticed that the measured acceptance dependence for the moments products of net protons present different behaviors, when compared with our calculations. For example,  $\sigma^{2}/M$ of net protons with an acceptance cut $0.8<p_T<2.0$ GeV is obviously below the Skellam base lines~\cite{Xu:2016qzd}, while our model calculations are above the Skellam baselines. For Au+Au collisions at $\sqrt{S_{\text{NN}}}$=7.7 GeV, the measured $\kappa\sigma^2$ presents an obvious centrality dependence, which dramatically deviate from the Skellam baseline in the most central collisions~\cite{Xu:2016qzd}. In contrast, our model calculations shows weak centrality dependence for  $\kappa\sigma^2$  at 7.7 GeV.
Again, our model simulations do not include all the possible effects as imprinted in nature, such as the conservation laws of net charges and net baryons, critical fluctuations, the spinodal instabilities of the first order phase transitions, etc., which are worthwhile to be further explored in the future.

Besides the acceptance cuts, different
centrality definitions also influence the volume fluctuations and the measured fluctuations of (net) charges and (net) protons. In Fig. 9, we calculate the moment products of net-charges and net-protons from {\tt iEBE-VISHNU} with different centrality cuts. In more details, the centrality bins are defined by the total charges within pseudo-rapidity cuts $0.5<|\eta|<1.0$ or $0.5<|\eta|<2.0$ (Fig.~9, left). For the case of net protons, the centralities are cut by the number of $\pi$, $K$ within $|\eta|<1.0$ or $|\eta|<2.0$ (Fig.~9, right).  In contrast to acceptance cut dependence shown in Fig.8, broader acceptance windows for the reference particles in the centrality cut lead to smaller deviations from the Skellam baselines here. As mentioned above, the volume fluctuations for Skewness and Kurtosis are dependent on  $(M_+-M_-)/(k+1)$. A border acceptance cut during the centrality definition leads to a lager value of $k$ of the reference particles, which reduces the related effects of volume fluctuations.
In Ref.~\cite{Luo:2011tp}, such effects are also mentioned as centrality resolution effects. In order to reduce the effects from volume fluctuations with a better centrality resolution, a larger acceptance cut for the reference particles in the centrality definition is preferred~\cite{Luo:2011tp}.

\section{Summary}\label{Summary and remarkable conclusions}

In this paper, we investigated the multiplicity fluctuations of (net) charges and (net) protons in Au+Au collisions at $\sqrt{s_{NN}} = 7.7$, $39$ and $200$ GeV, using the {\tt{iEBE-VISHNU}} hybrid model that combines 2+1-d viscous hydrodynamics with the {\tt UrQMD} hadron cascade model.  With a modified Cooper-Fryer freeze-out procedure,
the Poisson fluctuations have been added in the Monte-Carlo event generator that samples
thermal hadrons on the switching hypersurface between hydrodynamics and {\tt UrQMD}.  In our investigations, the moments/cumulants of (net) charges and (net) protons are calculated with the same centrality and acceptance cut as used in experiments, which also account various effects of initial state fluctuations,  volume fluctuations, hadronic scatterings and evolution, resonance decays, etc.

With well tuned parameters  that fit the mean values of pions, kaons and protons in Au+Au collisions at $\sqrt{s_{NN}} = 7.7$, $39$ and $200$ GeV, {\tt{iEBE-VISHNU}} roughly described the centrality dependent moments and cumulants of (net) charges and (net) protons measured in experiment. We also found that the {\tt{iEBE-VISHNU}} results are largely deviated from the Poisson/Skellam baselines for (net) charges, but are pretty close to the Poisson/Skellam baselines for (net) protons, especially at higher collision energy. Further comparison simulations have shown that the volume fluctuations play dominant roles to influence the higher moments of (net) charges, but do not significantly affect the multiplicity fluctuations of (net) protons. We also found that the effects from hadronic
evolutions are pretty small or even negligible for the multiplicity distributions of (net) charges and (net) protons. Considering that full {\tt UrQMD} hadronic evolution consumes a great portion of the calculation time in {\tt{iEBE-VISHNU}}, such finding may help to largely improve the numerical efficiency for the massive data simulations in the near future,
which makes it possible to calculate more realistic thermal fluctuation baselines with a realistic but simplified dynamical model.

Finally, we would like to emphasize that, although part of particle correlations have been included through the hadronic scatterings and resonance decays, our {\tt iEBE-VISHNU} calculations are still belong to framework of independent production since various thermal hadrons are independently emitted from the switching hyper-surface
according to the Cooper-Frye formula without further considering the conservation laws and other sources of correlations. Correspondingly, our calculations failed to nicely describe the standard variation $\sigma$ of net charges, which showed a certain gap between model and data. For an improved description of the data at higher collision energies and realistic predictions of the non-critical fluctuation baselines for the BES program, more effects, e.g. the conservation laws of net charges and net baryons~\cite{Bzdak:2012an}
, thermal fluctuations within hydrodynamics~\cite{Kapusta:2011gt}, improved initial state fluctuations, etc., should be further considered in our model calculations in the future.    \\

\textbf{Acknowledgments:} \\[-0.10in]

We thank the valuable discussions from V. Koch, U. Heinz, X. Luo and N. Xu.
This work is supported by the NSFC and the MOST under grant
Nos.11435001, 11675004 and 2015CB856900. H.~X. is partially supported by China Postdoctoral Science Foundation with
grant No.~2015M580908. We gratefully acknowledge the extensive computing resources provided to us
by Supercomputing Center of Chinese Academy of Science (SCCAS) and Tianhe-1A from the
National Supercomputing Center in Tianjin, China.


\begin{thebibliography}{}
	
%%%%%%%%%%%%%%%%%%%%%%%%%%%%%%%%%QCD phase goal%%%%%%%%%%%%%%%%%%%%%%%%%%%%%

\bibitem{Rev-Arsene:2004fa}
%{\bf BRAHMS} Collaboration,
  I.~Arsene {\it et al.} (BRAHMS Collaboration),
  %``Quark gluon plasma and color glass condensate at RHIC? The Perspective
  %  from the BRAHMS experiment,''
  Nucl.\ Phys.\  {\bf A757}, 1 (2005);
  %[nucl-ex/0410020].
%\bibitem{Back:2004je}
  B.~B.~Back {\it et al.} (PHOBOS Collaboration),
  %``The PHOBOS perspective on discoveries at RHIC,''
  %Nucl.\ Phys.\ {\bf A757}, 28 (2005)
  {\it ibid.,} p.\,28;
  %[nucl-ex/0410022].
%\bibitem{Adams:2005dq}
  J.~Adams {\it et al.} (STAR Collaboration),
  %``Experimental and theoretical challenges in the search for the quark
  %  gluon plasma: The STAR Collaboration's critical assessment of the
  %  evidence from RHIC collisions,''
  %Nucl.\ Phys.\ {\bf A757}, 102 (2005);
  {\it ibid.,} p.\,102;
  %[nucl-ex/0501009].
%\bibitem{Adcox:2004mh}
  K.~Adcox {\it et al.} (PHENIX Collaboration),
  %``Formation of dense partonic matter in relativistic nucleus-nucleus
  %  collisions at RHIC: Experimental evaluation by the PHENIX collaboration,''
  %Nucl.\ Phys.\ {\bf A757}, 184 (2005).
  {\it ibid.,} p.\,184.
  %[nucl-ex/0410003].

%\cite{Aoki:2006we}
\bibitem{Aoki:2006we}
  Y.~Aoki, G.~Endrodi, Z.~Fodor, S.~D.~Katz and K.~K.~Szabo,
  %``The Order of the quantum chromodynamics transition predicted by the standard model of particle physics,''
  Nature {\bf 443}, 675 (2006).
 % doi:10.1038/nature05120
 % [hep-lat/0611014].
  %%CITATION = doi:10.1038/nature05120;%%
  %961 citations counted in INSPIRE as of 30 Jun 2017

\bibitem{Aggarwal:2010cw} M.~M.~Aggarwal \textit{et al.} [STAR Collaboration],
%``An Experimental Exploration of the QCD Phase Diagram: The Search for the Critical Point and the Onset of De-confinement,''
arXiv:1007.2613. % [nucl-ex].

%\cite{Akiba:2015jwa}
\bibitem{Akiba:2015jwa}
Y.~Akiba {\it et al.},
%``The Hot QCD White Paper: Exploring the Phases of QCD at RHIC and the LHC,''
arXiv:1502.02730 [nucl-ex].
%%CITATION = ARXIV:1502.02730;%%
%41 citations counted in INSPIRE as of 16 Jun 2017


%%%%%%%%%%%%%%%%%%%%%review papers%%%%%%%%%%%%%%%%%%%%%%%%%%%%%%%%%%
%\cite{Stephanov:2004wx}
\bibitem{Stephanov:2004wx}
M.~A.~Stephanov,
%``QCD phase diagram and the critical point,''
Prog.\ Theor.\ Phys.\ Suppl.\  {\bf 153}, 139 (2004);
[Int.\ J.\ Mod.\ Phys.\ A {\bf 20}, 4387 (2005)].
%doi:10.1142/S0217751X05027965
%[hep-ph/0402115].
%%CITATION = doi:10.1142/S0217751X05027965;%%
%305 citations counted in INSPIRE as of 07 Apr 2016




  %\cite{Asakawa:2015ybt}
\bibitem{Asakawa:2015ybt}
  M.~Asakawa and M.~Kitazawa,
  %``Fluctuations of conserved charges in relativistic heavy ion collisions: An introduction,''
  Prog.\ Part.\ Nucl.\ Phys.\  {\bf 90}, 299 (2016).
  %doi:10.1016/j.ppnp.2016.04.002
  %[arXiv:1512.05038 [nucl-th]].
  %%CITATION = doi:10.1016/j.ppnp.2016.04.002;%%
  %20 citations counted in INSPIRE as of 16 Jun 2017

  %\cite{Luo:2017faz}
\bibitem{Luo:2017faz}
  X.~Luo and N.~Xu,
  %``Search for the QCD Critical Point with Fluctuations of Conserved Quantities in Relativistic Heavy-Ion Collisions at RHIC : An Overview,''
  arXiv:1701.02105 [nucl-ex].
  %%CITATION = ARXIV:1701.02105;%%
  %17 citations counted in INSPIRE as of 16 Jun 2017

%%%%%%%%%%%%%%%%%%%%%%%%%%%%%%%High Moments%%%%%%%%%%%%%%%%%%%%%%%%%%%%%%%%%%%
 %\cite{Stephanov:2008qz}
  \bibitem{Stephanov:2008qz}
  M.~A.~Stephanov,
  %``Non-Gaussian fluctuations near the QCD critical point,''
  Phys.\ Rev.\ Lett.\  {\bf 102} 032301 (2009).
  %doi:10.1103/PhysRevLett.102.032301
  %[arXiv:0809.3450 [hep-ph]].
  %%CITATION = doi:10.1103/PhysRevLett.102.032301;%%
  %254 citations counted in INSPIRE as of 07 Apr 2016


  %\cite{Athanasiou:2010kw}
  \bibitem{Athanasiou:2010kw}
  C.~Athanasiou, K.~Rajagopal and M.~Stephanov,
  %``Using Higher Moments of Fluctuations and their Ratios in the Search for the QCD Critical Point,''
  Phys.\ Rev.\ D {\bf 82}, 074008 (2010).
  %doi:10.1103/PhysRevD.82.074008
  %[arXiv:1006.4636 [hep-ph]].
  %%CITATION = doi:10.1103/PhysRevD.82.074008;%%
  %83 citations counted in INSPIRE as of 07 Apr 2016




%%%%%%%%%%%%%%%%%%%%%%%%%%%%%%%%%%%%%%%%%%%%%%%%%%%%%%%%%%%%%%%%%%%%%%%%%%%%%%

%%%%%%%%%%%%%%%%%%%expriments paper%%%%%%%%%%%%%%%%%%%%%%%%%%%%%%%%%%%%%
%\cite{Aggarwal:2010wy}
\bibitem{Aggarwal:2010wy}
M.~M.~Aggarwal {\it et al.} [STAR Collaboration],
%``Higher Moments of Net-proton Multiplicity Distributions at RHIC,''
Phys.\ Rev.\ Lett.\  {\bf 105}, 022302 (2010).
%doi:10.1103/PhysRevLett.105.022302
%[arXiv:1004.4959 [nucl-ex]].
%%CITATION = doi:10.1103/PhysRevLett.105.022302;%%
%176 citations counted in INSPIRE as of 07 Apr 2016


%\cite{Adamczyk:2013dal}
\bibitem{Adamczyk:2013dal}
L.~Adamczyk {\it et al.} [STAR Collaboration],
%``Energy Dependence of Moments of Net-proton Multiplicity Distributions at RHIC,''
Phys.\ Rev.\ Lett.\  {\bf 112}, 032302 (2014).
%doi:10.1103/PhysRevLett.112.032302
%[arXiv:1309.5681 [nucl-ex]].
%%CITATION = doi:10.1103/PhysRevLett.112.032302;%%
%133 citations counted in INSPIRE as of 07 Apr 2016

%\cite{Adamczyk:2014fia}
\bibitem{Adamczyk:2014fia}
L.~Adamczyk {\it et al.} [STAR Collaboration],
%``Beam energy dependence of moments of the net-charge multiplicity distributions in Au+Au collisions at RHIC,''
Phys.\ Rev.\ Lett.\  {\bf 113}, 092301 (2014).
%doi:10.1103/PhysRevLett.113.092301
%[arXiv:1402.1558 [nucl-ex]].
%%CITATION = doi:10.1103/PhysRevLett.113.092301;%%
%93 citations counted in INSPIRE as of 07 Apr 2016

%\cite{Luo:2015ewa}
\bibitem{Luo:2015ewa}
X.~Luo [STAR Collaboration],
%``Energy Dependence of Moments of Net-Proton and Net-Charge Multiplicity Distributions at STAR,''
PoS CPOD {\bf 2014}, 019 (2015).
%[arXiv:1503.02558 [nucl-ex]].
%%CITATION = ARXIV:1503.02558;%%
%58 citations counted in INSPIRE as of 16 Jun 2017
%%%%%%%%%%%%%%%%%%%%%%%%%%%%%%%%%%%%%%%%%%%%%%%%%%%%%%%%%%%%%%%%%%%%%%%%%%%%%%
\bibitem{Begun:2004gs}
  V.~V.~Begun, M.~Gazdzicki, M.~I.~Gorenstein and O.~S.~Zozulya,
  %``Particle number fluctuations in canonical ensemble,''
  Phys.\ Rev.\ C {\bf 70}, 034901 (2004).
  %doi:10.1103/PhysRevC.70.034901
  %[nucl-th/0404056].
  %%CITATION = doi:10.1103/PhysRevC.70.034901;%%
  %115 citations counted in INSPIRE as of 30 Jun 2017
%%%%%%%%%%%%%%%%%%%%%%%%%%%%%%%%%%%%%%%%%%%%%%%%%%%%%%%%%%%%%%%%%%%%%%%%%%%%%
%\cite{Cleymans:2004iu}
\bibitem{Cleymans:2004iu}
  J.~Cleymans, K.~Redlich and L.~Turko,
  %``Probability distributions in statistical ensembles with conserved charges,''
  Phys.\ Rev.\ C {\bf 71}, 047902 (2005).
  %doi:10.1103/PhysRevC.71.047902
  %[hep-th/0412262].
  %%CITATION = doi:10.1103/PhysRevC.71.047902;%%
  %37 citations counted in INSPIRE as of 30 Jun 2017

%\cite{Landau:1980mil}
\bibitem{Landau:1980mil}
  L.~D.~Landau and E.~M.~Lifshitz,
  ``Statistical Physics, Part 1''.


%%%%%%%%%%%%%%%%%%%%%%%imporantce of thermal fluctuation %%%%%%%%%%%%%%%%%%%%%
 %\cite{Karsch:2010ck}
\bibitem{Karsch:2010ck}
  F.~Karsch and K.~Redlich,
  %``Probing freeze-out conditions in heavy ion collisions with moments of charge fluctuations,''
  Phys.\ Lett.\ B {\bf 695}, 136 (2011).
  %doi:10.1016/j.physletb.2010.10.046
  %[arXiv:1007.2581 [hep-ph]].
  %%CITATION = doi:10.1016/j.physletb.2010.10.046;%%
  %191 citations counted in INSPIRE as of 30 Jun 2017

  %\cite{Garg:2013ata}
\bibitem{Garg:2013ata}
  P.~Garg, D.~K.~Mishra, P.~K.~Netrakanti, B.~Mohanty, A.~K.~Mohanty, B.~K.~Singh and N.~Xu,
  %``Conserved number fluctuations in a hadron resonance gas model,''
  Phys.\ Lett.\ B {\bf 726}, 691 (2013).
  %doi:10.1016/j.physletb.2013.09.019
  %[arXiv:1304.7133 [nucl-ex]].
  %%CITATION = doi:10.1016/j.physletb.2013.09.019;%%
  %62 citations counted in INSPIRE as of 30 Jun 2017



%\cite{Begun:2004gs}




  %\cite{Begun:2006uu}
\bibitem{Begun:2006uu}
  V.~V.~Begun, M.~Gazdzicki, M.~I.~Gorenstein, M.~Hauer, V.~P.~Konchakovski and B.~Lungwitz,
  %``Multiplicity fluctuations in relativistic nuclear collisions: Statistical model versus experimental data,''
  Phys.\ Rev.\ C {\bf 76}, 024902 (2007).
  %doi:10.1103/PhysRevC.76.024902
  %[nucl-th/0611075].
  %%CITATION = doi:10.1103/PhysRevC.76.024902;%%
  %64 citations counted in INSPIRE as of 30 Jun 2017

  %\cite{Chatterjee:2009km}
\bibitem{Chatterjee:2009km}
  S.~Chatterjee, R.~M.~Godbole and S.~Gupta,
  %``Stabilizing hadron resonance gas models,''
  Phys.\ Rev.\ C {\bf 81}, 044907 (2010).
  %doi:10.1103/PhysRevC.81.044907
  %[arXiv:0906.2523 [hep-ph]].
  %%CITATION = doi:10.1103/PhysRevC.81.044907;%%
  %33 citations counted in INSPIRE as of 30 Jun 2017





 %\cite{Alba:2014eba}
\bibitem{Alba:2014eba}
  P.~Alba, W.~Alberico, R.~Bellwied, M.~Bluhm, V.~Mantovani Sarti, M.~Nahrgang and C.~Ratti,
  %``Freeze-out conditions from net-proton and net-charge fluctuations at RHIC,''
  Phys.\ Lett.\ B {\bf 738}, 305 (2014).
  %doi:10.1016/j.physletb.2014.09.052
  %[arXiv:1403.4903 [hep-ph]].
  %%CITATION = doi:10.1016/j.physletb.2014.09.052;%%
  %87 citations counted in INSPIRE as of 30 Jun 2017



  %\cite{Bhattacharyya:2015zka}
\bibitem{Bhattacharyya:2015zka}
  A.~Bhattacharyya, R.~Ray, S.~Samanta and S.~Sur,
  %``Thermodynamics and fluctuations of conserved charges in a hadron resonance gas model in a finite volume,''
  Phys.\ Rev.\ C {\bf 91}, no. 4, 041901 (2015).
  %doi:10.1103/PhysRevC.91.041901
  %[arXiv:1502.00889 [hep-ph]].
  %%CITATION = doi:1:0.1103/PhysRevC.91.041901;%%
  %20 citations counted in INSPIRE as of 30 Jun 2017

  %\cite{BraunMunzinger:2011dn}
\bibitem{BraunMunzinger:2011dn}
  P.~Braun-Munzinger, B.~Friman, F.~Karsch, K.~Redlich and V.~Skokov,
  %``Net-proton probability distribution in heavy ion collisions,''
  Phys.\ Rev.\ C {\bf 84}, 064911 (2011).
  %doi:10.1103/PhysRevC.84.064911
  %[arXiv:1107.4267 [hep-ph]].
  %%CITATION = doi:10.1103/PhysRevC.84.064911;%%
  %37 citations counted in INSPIRE as of 30 Jun 2017

  %\cite{Mishra:2016qyj}
\bibitem{Mishra:2016qyj}
  D.~K.~Mishra, P.~Garg, P.~K.~Netrakanti and A.~K.~Mohanty,
  %``Effect of resonance decay on conserved number fluctuations in a hadron resonance gas model,''
  Phys.\ Rev.\ C {\bf 94}, no. 1, 014905 (2016).
  %doi:10.1103/PhysRevC.94.014905
  %[arXiv:1607.01875 [hep-ph]].
  %%CITATION = doi:10.1103/PhysRevC.94.014905;%%
  %8 citations counted in INSPIRE as of 30 Jun 2017




%\cite{Nahrgang:2014fza}
\bibitem{Nahrgang:2014fza}
M.~Nahrgang, M.~Bluhm, P.~Alba, R.~Bellwied and C.~Ratti,
%``Impact of resonance regeneration and decay on the net-proton fluctuations in a hadron resonance gas,''
Eur.\ Phys.\ J.\ C {\bf 75}, no. 12, 573 (2015).
%doi:10.1140/epjc/s10052-015-3775-0
%[arXiv:1402.1238 [hep-ph]].
%%CITATION = doi:10.1140/epjc/s10052-015-3775-0;%%
%44 citations counted in INSPIRE as of 16 Jun 2017
%\cite{Luo:2014tga}




%\cite{Bazavov:2012jq}
\bibitem{Bazavov:2012jq}
  A.~Bazavov {\it et al.} [HotQCD Collaboration],
  %``Fluctuations and Correlations of net baryon number, electric charge, and strangeness: A comparison of lattice QCD results with the hadron resonance gas model,''
  Phys.\ Rev.\ D {\bf 86}, 034509 (2012).




%\cite{Borsanyi:2013hza}
\bibitem{Borsanyi:2013hza}
  S.~Borsanyi, Z.~Fodor, S.~D.~Katz, S.~Krieg, C.~Ratti and K.~K.~Szabo,
  %``Freeze-out parameters: lattice meets experiment,''
  Phys.\ Rev.\ Lett.\  {\bf 111}, 062005 (2013).





  %\cite{Fu:2013gga}
\bibitem{Fu:2013gga}
  J.~Fu,
  %``Higher moments of net-proton multiplicity distributions in heavy ion collisions at chemical freeze-out,''
  Phys.\ Lett.\ B {\bf 722}, 144 (2013).
  %doi:10.1016/j.physletb.2013.04.018
  %%CITATION = doi:10.1016/j.physletb.2013.04.018;%%
  %31 citations counted in INSPIRE as of 30 Jun 2017


\bibitem{Luo:2014tga}
X.~Luo, B.~Mohanty and N.~Xu,
%``Baseline for the cumulants of net-proton distributions at STAR,''
Nucl.\ Phys.\ A {\bf 931}, 808 (2014).
%doi:10.1016/j.nuclphysa.2014.08.105
%[arXiv:1408.0495 [nucl-ex]].
%%CITATION = doi:10.1016/j.nuclphysa.2014.08.105;%%
%13 citations counted in INSPIRE as of 16 Jun 2017

\bibitem{Netrakanti:2014mta}
P.~K.~Netrakanti, X.~F.~Luo, D.~K.~Mishra, B.~Mohanty, A.~Mohanty and N.~Xu,
%``Baseline measures for net-proton distributions in high energy heavy-ion collisions,''
Nucl.\ Phys.\ A {\bf 947}, 248 (2016).



  %%CITATION = INSPIRE-1411177;%%


%%%%%%%%%%%%%%%%%%%%%%%%%%%%%HRG Model%%%%%%%%%%%%%%%%%%%%%%%%%%%%%%%%%%%%%%%
%\cite{BraunMunzinger:2003zd}
%%%%%%%%%%%%%%%%%%%%%%%%%%%%%%%%%%%%%%%%%%%%%%%%%%%%%%%%%%%%%%%%%%%%%%%%%%%%%%

%\cite{Jiang:2015hri}
\bibitem{Jiang:2015hri}
  L.~Jiang, P.~Li and H.~Song,
  %``Correlated fluctuations near the QCD critical point,''
  Phys.\ Rev.\ C {\bf 94}, no. 2, 024918 (2016);
%
%\cite{Jiang:2015cnt}
%\bibitem{Jiang:2015cnt}
%  L.~Jiang, P.~Li and H.~Song,
  %``Multiplicity fluctuations of net protons on the hydrodynamic freeze-out surface,''
  Nucl.\ Phys.\ A {\bf 956}, 360 (2016).

\bibitem{Jiang:2017mji}
	  L.~Jiang, S.~Wu and H.~Song,
	    %``Dynamical fluctuations in critical regime and across the 1st order phase transition,''
		   arXiv:1704.04765 [nucl-th].
			  %%CITATION = ARXIV:1704.04765;%%

%%%%%%%%%%%%%%%%%%%%%%%%Volume Effects%%%%%%%%%%%%%%%%%%%%%%%%%%%%%%%%%%%%%%%%%%%


  %\cite{Jeon:2003gk}
\bibitem{Jeon:2003gk}
  S.~Jeon and V.~Koch,
  %``Event by event fluctuations,''
  In *Hwa, R.C. (ed.) et al.: Quark gluon plasma* 430-490.
 % [hep-ph/0304012].
  %%CITATION = HEP-PH/0304012;%%
  %125 citations counted in INSPIRE as of 13 Jun 2016


%\cite{Skokov:2012ds}
\bibitem{Skokov:2012ds}
  V.~Skokov, B.~Friman and K.~Redlich,
  %``Volume Fluctuations and Higher Order Cumulants of the Net Baryon Number,''
  Phys.\ Rev.\ C {\bf 88}, 034911 (2013).
  %doi:10.1103/PhysRevC.88.034911
  %[arXiv:1205.4756 [hep-ph]].
  %%CITATION = doi:10.1103/PhysRevC.88.034911;%%
  %26 citations counted in INSPIRE as of 13 Jun 2016


  %\cite{Xu:2016qzd}
\bibitem{Xu:2016qzd}
	H.~j.~Xu,
	%``Cumulants of multiplicity distributions in most-central heavy-ion collisions,''
	Phys.\ Rev.\ C {\bf 94}, no. 5, 054903 (2016).
	%doi:10.1103/PhysRevC.94.054903
	%[arXiv:1602.07089 [nucl-th]].
	%%CITATION = doi:10.1103/PhysRevC.94.054903;%%
	%6 citations counted in INSPIRE as of 23 Jun 2017


%\cite{Xu:2016skm}
\bibitem{Xu:2016skm}
H.~j.~Xu,
%``Effects of volume corrections and resonance decays on cumulants of net-charge distributions in a Monte Carlo hadron resonance gas model,''
Phys.\ Lett.\ B {\bf 765}, 188 (2017);
%doi:10.1016/j.physletb.2016.12.015
%[arXiv:1612.06485 [nucl-th]].
%%CITATION = doi:10.1016/j.physletb.2016.12.015;%%
%2 citations counted in INSPIRE as of 07 May 2017
%\cite{Xu:2016csh}
%\bibitem{Xu:2016csh}
%H.~j.~Xu,
%``Importance of volume corrections on the net-charge distributions at the RHIC BES energies,''
arXiv:1610.08591 [nucl-th].
%%CITATION = ARXIV:1610.08591;%%
%2 citations counted in INSPIRE as of 07 May 2017

%%%%%%%%%%%%%%%%%%%%%%%%%%%%%%%%%%%%%%%%%%%%%%%%%%%%%%%%%%%%%%%%%%%%%%%%%%%%%%%%%%%%


%%%%%%%%%%%%%%%%%%%%%%%%%%Acceptance and efficiency%%%%%%%%%%%%%%%%%%%%%%%%%%%%%%%%
%\cite{Bzdak:2012ab}
\bibitem{Bzdak:2012ab}
A.~Bzdak and V.~Koch,
%``Acceptance corrections to net baryon and net charge cumulants,''
Phys.\ Rev.\ C {\bf 86}, 044904 (2012).
%doi:10.1103/PhysRevC.86.044904
%[arXiv:1206.4286 [nucl-th]].
%%CITATION = doi:10.1103/PhysRevC.86.044904;%%
%70 citations counted in INSPIRE as of 16 Jun 2017

%\cite{Karsch:2015zna}
\bibitem{Karsch:2015zna}
F.~Karsch, K.~Morita and K.~Redlich,
%``Effects of kinematic cuts on net-electric charge fluctuations,''
Phys.\ Rev.\ C {\bf 93}, no. 3, 034907 (2016).
%doi:10.1103/PhysRevC.93.034907
%[arXiv:1508.02614 [hep-ph]].
%%CITATION = doi:10.1103/PhysRevC.93.034907;%%
%7 citations counted in INSPIRE as of 07 Apr 2016


%\cite{Petersen:2015pcy}
\bibitem{Petersen:2015pcy}
H.~Petersen, D.~Oliinychenko, J.~Steinheimer and M.~Bleicher,
%``Influence of kinematic cuts on the net charge distribution,''
arXiv:1512.05603 [nucl-th].
%%CITATION = ARXIV:1512.05603;%%

%\cite{Ling:2015yau}
\bibitem{Ling:2015yau}
B.~Ling and M.~A.~Stephanov,
%``Acceptance dependence of fluctuation measures near the QCD critical point,''
Phys.\ Rev.\ C {\bf 93}, no. 3, 034915 (2016).
% doi:10.1103/PhysRevC.93.034915
%[arXiv:1512.09125 [nucl-th]].
%%CITATION = doi:10.1103/PhysRevC.93.034915;%%
%4 citations counted in INSPIRE as of 07 Jul 2016


%%%%%%%%%%%%%%%%%%%%%%%%%%%%%%%%%%%%%%conservation law%%%%%%%%%%%%%%%%%%%%%%%%%%%%
%\cite{Jeon:2000wg}
\bibitem{Jeon:2000wg}
S.~Jeon and V.~Koch,
%``Charged particle ratio fluctuation as a signal for QGP,''
Phys.\ Rev.\ Lett.\  {\bf 85}, 2076 (2000).
%doi:10.1103/PhysRevLett.85.2076
%[hep-ph/0003168].
%%CITATION = doi:10.1103/PhysRevLett.85.2076;%%
%386 citations counted in INSPIRE as of 16 Jun 2017

%\cite{Bzdak:2012an}
\bibitem{Bzdak:2012an}
  A.~Bzdak, V.~Koch and V.~Skokov,
  %``Baryon number conservation and the cumulants of the net proton distribution,''
  Phys.\ Rev.\ C {\bf 87}, no. 1, 014901 (2013).
  %doi:10.1103/PhysRevC.87.014901
  %[arXiv:1203.4529 [hep-ph]].
  %%CITATION = doi:10.1103/PhysRevC.87.014901;%%
  %49 citations counted in INSPIRE as of 07 Apr 2016

%\cite{Sakaida:2014pya}
\bibitem{Sakaida:2014pya}
  M.~Sakaida, M.~Asakawa and M.~Kitazawa,
  %``Effects of global charge conservation on time evolution of cumulants of conserved charges in relativistic heavy ion collisions,''
  Phys.\ Rev.\ C {\bf 90}, no. 6, 064911 (2014).
  %doi:10.1103/PhysRevC.90.064911
  %[arXiv:1409.6866 [nucl-th]].
  %%CITATION = doi:10.1103/PhysRevC.90.064911;%%
  %14 citations counted in INSPIRE as of 07 Apr 2016

%%%%%%%%%%%%%%%%%%%%%%%Resonence decay effects%%%%%%%%%%%%%%%%%%%%%%%%

%\cite{Kitazawa:2011wh}
\bibitem{Kitazawa:2011wh}
M.~Kitazawa and M.~Asakawa,
%``Revealing baryon number fluctuations from proton number fluctuations in relativistic heavy ion collisions,''
Phys.\ Rev.\ C {\bf 85}, 021901 (2012).
%doi:10.1103/PhysRevC.85.021901
%[arXiv:1107.2755 [nucl-th]].
%%CITATION = doi:10.1103/PhysRevC.85.021901;%%
%70 citations counted in INSPIRE as of 16 Jun 2017


%\cite{Kitazawa:2012at}
\bibitem{Kitazawa:2012at}
M.~Kitazawa and M.~Asakawa,
%``Relation between baryon number fluctuations and experimentally observed proton number fluctuations in relativistic heavy ion collisions,''
Phys.\ Rev.\ C {\bf 86}, 024904 (2012)
Erratum: [Phys.\ Rev.\ C {\bf 86}, 069902 (2012)].
%doi:10.1103/PhysRevC.86.024904, 10.1103/PhysRevC.86.069902
%[arXiv:1205.3292 [nucl-th]].
%%CITATION = doi:10.1103/PhysRevC.86.024904, 10.1103/PhysRevC.86.069902;%%
%72 citations counted in INSPIRE as of 16 Jun 2017


%=== chemical thermal equilibrium
%\cite{Hirano:2005xf}
\bibitem{Hirano:2005xf}
  T.~Hirano, U.~W.~Heinz, D.~Kharzeev, R.~Lacey and Y.~Nara,
  %``Hadronic dissipative effects on elliptic flow in ultrarelativistic heavy-ion collisions,''
  Phys.\ Lett.\ B {\bf 636}, 299 (2006);
%\bibitem{Hirano:2005wx}
  T.~Hirano and M.~Gyulassy,
  %``Perfect fluidity of the quark gluon plasma core as seen through its dissipative hadronic corona,''
  Nucl.\ Phys.\ A {\bf 769}, 71 (2006).

\bibitem{Teaney:2002aj}
  D.~Teaney,
  %``Chemical freezeout in heavy ion collisions,''
  arXiv:nucl-th/0204023.
  %%CITATION = NUCL-TH/0204023;%%

\bibitem{Hirano:2002ds}
  T.~Hirano and K.~Tsuda,
  %``Collective flow and two pion correlations from a relativistic hydrodynamic
  %model with early chemical freeze out,''
  Phys.\ Rev.\  C {\bf 66}, 054905 (2002).
  %[arXiv:nucl-th/0205043].
  %%CITATION = PHRVA,C66,054905;%%

\bibitem{Kolb:2002ve}
  P.~F.~Kolb and R.~Rapp,
  %``Transverse flow and hadro-chemistry in Au+Au collisions at s(NN)**(1/2) =
  %200-GeV,''
  Phys.\ Rev.\  C {\bf 67}, 044903 (2003).
  %[arXiv:hep-ph/0210222].
  %%CITATION = PHRVA,C67,044903;%%

\bibitem{Huovinen:2007xh}
  P.~Huovinen,
  %``Chemical freeze-out temperature in hydrodynamical description of Au+Au
  %collisions at sqrt(s_NN) = 200 GeV,''
  Eur.\ Phys.\ J.\  A {\bf 37}, 121 (2008).
  %[arXiv:0710.4379 [nucl-th]].
  %%CITATION = EPHJA,A37,121;%%

%\cite{Xu:2016qjd}
\bibitem{Xu:2016qjd}
  J.~Xu, S.~Yu, F.~Liu and X.~Luo,
  %``Cumulants of net-proton, net-kaon, and net-charge multiplicity distributions in Au + Au collisions at $\sqrt {s_{NN}}$=7.7 , 11.5, 19.6, 27, 39, 62.4, and 200 GeV within the UrQMD model,''
  Phys.\ Rev.\ C {\bf 94}, no. 2, 024901 (2016).
 % doi:10.1103/PhysRevC.94.024901
  %[arXiv:1606.03900 [nucl-ex]].
  %%CITATION = doi:10.1103/PhysRevC.94.024901;%%
  %8 citations counted in INSPIRE as of 30 Jun 2017

%\cite{He:2017zpg}
\bibitem{He:2017zpg}
  S.~He and X.~Luo,
  %``Proton Cumulants and Correlation Functions in Au + Au Collisions at $\sqrt{s_\mathrm{NN}}$=7.7-200 GeV from UrQMD Model,''
  arXiv:1704.00423 [nucl-ex].


%\cite{Yang:2016xga}
\bibitem{Yang:2016xga}
  Z.~Yang, X.~Luo and B.~Mohanty,
  %``Baryon-Strangeness Correlations in Au+Au Collisions at $\sqrt{s_\mathrm{NN}}$=7.7-200 GeV from the UrQMD model,''
  Phys.\ Rev.\ C {\bf 95}, no. 1, 014914 (2017).
 % doi:10.1103/PhysRevC.95.014914
 % [arXiv:1610.07580 [nucl-ex]].
  %%CITATION = doi:10.1103/PhysRevC.95.014914;%%
  %1 citations counted in INSPIRE as of 23 Jul 2017

%\cite{Zhou:2017jfk}
\bibitem{Zhou:2017jfk}
  C.~Zhou, J.~Xu, X.~Luo and F.~Liu,
  %``Cumulants of event-by-event net-strangeness distributions in Au+Au collisions at $\sqrt{s_\mathrm{NN}}$=7.7-200 GeV from UrQMD model,''
  Phys.\ Rev.\ C {\bf 96}, no. 1, 014909 (2017).
  %doi:10.1103/PhysRevC.96.014909
  %[arXiv:1703.09114 [nucl-ex]].
  %%CITATION = doi:10.1103/PhysRevC.96.014909;%%
  %2 citations counted in INSPIRE as of 23 Jul 2017


%\cite{He:2016uei}
\bibitem{He:2016uei}
  S.~He, X.~Luo, Y.~Nara, S.~Esumi and N.~Xu,
  %``Effects of Nuclear Potential on the Cumulants of Net-Proton and Net-Baryon Multiplicity Distributions in Au+Au Collisions at $\sqrt{s_{\text{NN}}} = 5\,\text{GeV}$,''
  Phys.\ Lett.\ B {\bf 762}, 296 (2016).




%%%%%%%%%%%%%%%%%%%%%IEBE-VISHNU%%%%%%%%%%%%%%%%%%%%%%%%%%%%%%%%%%%%%%%%%%%%%%%%%%%%%%%%%%
%
 %\cite{Shen:2014vra}
 \bibitem{Shen:2014vra}
 C.~Shen, Z.~Qiu, H.~Song, J.~Bernhard, S.~Bass and U.~Heinz,
 %``The iEBE-VISHNU code package for relativistic heavy-ion collisions,''
 Comput.\ Phys.\ Commun.\  {\bf 199}, 61 (2016).
 %[arXiv:1409.8164 %[nucl-th]].


\bibitem{Song:2010aq}
  H.~Song, S.~A.~Bass and U.~Heinz,
  %``Viscous QCD matter in a hybrid hydrodynamic+Boltzmann approach,''
  Phys.\ Rev.\  C {\bf 83}, 024912 (2011).

%=============MC-Glb
\bibitem{Miller:2007ri}
  M.~L.~Miller, K.~Reygers, S.~J.~Sanders and P.~Steinberg,
  %``Glauber modeling in high energy nuclear collisions,''
  Ann.\ Rev.\ Nucl.\ Part.\ Sci.\  {\bf 57}, 205 (2007).

%=============MC-KLN============
\bibitem{Drescher:2006ca}
  H.~J.~Drescher and Y.~Nara,
  %``Effects of fluctuations on the initial eccentricity from the color  glass
  %condensate in heavy ion collisions,''
  Phys.\ Rev.\  C {\bf 75}, 034905 (2007);
  %[arXiv:nucl-th/0611017].
  %%CITATION = PHRVA,C75,034905;%%
%\bibitem{Drescher:2007ax}
  %H.~J.~Drescher and Y.~Nara,
  %``Eccentricity fluctuations from the Color Glass Condensate at RHIC and
  %LHC,''
  % Phys.\ Rev.\  C
   Phys.\ Rev.\  C {\bf 76}, 041903(R) (2007).
  %[arXiv:0707.0249 [nucl-th]].
  %%CITATION = PHRVA,C76,041903;%%
%
\bibitem{Hirano:2009ah}
  T.~Hirano and Y.~Nara,
  %``Eccentricity fluctuation effects on elliptic flow in relativistic
  %  heavy ion collisions,''
  Phys.\ Rev.\  C {\bf 79}, 064904 (2009);
  %[arXiv:0904.4080 [nucl-th]].
  %%CITATION = PHRVA,C79,064904;%%
%\bibitem{Hirano:2010jg}
  T.~Hirano, P.~Huovinen and Y.~Nara,
  %``Elliptic flow in U+U collisions at $\sqrt{s_{NN}}=200$ GeV and in Pb--Pb collisions at $\sqrt{s_{NN}}=2.76$ TeV: Prediction from a hybrid approach,''
  Phys.\ Rev.\ C {\bf 83}, 021902 (2011).


%
%\cite{Moreland:2014oya}
\bibitem{Moreland:2014oya}
  J.~S.~Moreland, J.~E.~Bernhard and S.~A.~Bass,
  %``Alternative ansatz to wounded nucleon and binary collision scaling in high-energy nuclear collisions,''
  Phys.\ Rev.\ C {\bf 92}, no. 1, 011901 (2015).

%\cite{Xu:2016hmp}
\bibitem{Xu:2016hmp}
  H.~j.~Xu, Z.~Li and H.~Song,
  %``High-order flow harmonics of identified hadrons in 2.76A TeV Pb + Pb collisions,''
  Phys.\ Rev.\ C {\bf 93}, no. 6, 064905 (2016).
 % doi:10.1103/PhysRevC.93.064905
 % [arXiv:1602.02029 [nucl-th]].


\bibitem{Song:2007fn}
  H.~Song and U.~Heinz,
  %``Suppression of elliptic flow in a minimally viscous quark-gluon plasma,''
  Phys.\ Lett.\  {\bf B658}, 279 (2008);
  %[arXiv:0709.0742 [nucl-th]].
%\bibitem{Song:2007ux}
  %H.~Song and U.~Heinz,
  %``Causal viscous hydrodynamics in 2+1 dimensions for relativistic
  %  heavy-ion collisions,''
  Phys.\ Rev.\  C {\bf 77}, 064901 (2008);
  %[arXiv:0712.3715 [nucl-th]].
%\bibitem{Song:2008si}
  %H.~Song and U.~Heinz,
  %``Multiplicity scaling in ideal and viscous hydrodynamics,''
  Phys.\ Rev.\ C {\bf 78}, 024902 (2008).
  %[arXiv:0805.1756 [nucl-th]].
%\cite{Cooper:1974mv}

\bibitem{Song:2009gc}
  H.~Song, Ph.D Thesis, The Ohio State University (August 2009),
  %``Causal Viscous Hydrodynamics for Relativistic Heavy Ion Collisions,''
  arXiv:0908.3656 [nucl-th].
  %%CITATION = ARXIV:0908.3656;%%

%%%%%%%%%%%%%%%%%%%%%%%%%%%%%%%%%URQMD and AMPT　Model%%%%%%%%%%%%%%%%%%%%
%-========URQMD==== \cite{Bass:1998ca,Bleicher:1999xi}
%\cite{Bass:1998ca}
 \bibitem{Bass:1998ca}
  S.~A.~Bass {\it et al.},
  %``Microscopic models for ultrarelativistic heavy ion collisions,''
  Prog.\ Part.\ Nucl.\ Phys.\  {\bf 41}, 255 (1998).
  %[arXiv:nucl-th/9803035].
  %%CITATION = PPNPD,41,225;%%

%\cite{Bleicher:1999xi}
\bibitem{Bleicher:1999xi}
M.~Bleicher {\it et al.},
%``Relativistic hadron hadron collisions in the ultrarelativistic quantum molecular dynamics model,''
J.\ Phys.\ G {\bf 25}, 1859 (1999).
%doi:10.1088/0954-3899/25/9/308
%[hep-ph/9909407].
%%CITATION = doi:10.1088/0954-3899/25/9/308;%%
%844 citations counted in INSPIRE as of 16 Jun 2017

\bibitem{ShenPhD}
  C.~Shen, Ph.D Thesis, The Ohio State University (August 2014).
  %``Causal Viscous Hydrodynamics for Relativistic Heavy Ion Collisions,''


%\cite{Song:2017wtw}
\bibitem{Song:2017wtw}
  H.~Song, Y.~Zhou and K.~Gajdosova,
  %``Collective flow and hydrodynamics in large and small systems at the LHC,''
  Nucl.\ Sci.\ Tech.\  {\bf 28}, no. 7, 99 (2017).




\bibitem{Cooper:1974mv}
  F.~Cooper and G.~Frye,
  %``Comment on the Single Particle Distribution in the Hydrodynamic and Statistical Thermodynamic Models of Multiparticle Production,''
  Phys.\ Rev.\ D {\bf 10}, 186 (1974).
  %doi:10.1103/PhysRevD.10.186
  %%CITATION = doi:10.1103/PhysRevD.10.186;%%
  %654 citations counted in INSPIRE as of 07 Apr 2016


			%\cite{Abelev:2008ab}
\bibitem{Abelev:2008ab}
B.~I.~Abelev {\it et al.} [STAR Collaboration],
%``Systematic Measurements of Identified Particle Spectra in $p p, d^+$ Au and Au+Au Collisions from STAR,''
  Phys.\ Rev.\ C {\bf 79}, 034909 (2009).



\bibitem{Song:2013qma}
  H.~Song, S.~Bass and U.~W.~Heinz,
  %``Spectra and elliptic flow for identified hadrons in 2.76A TeV Pb + Pb collisions,''
  Phys.\ Rev.\ C {\bf 89}, no. 3, 034919 (2014); X.~Zhu, F.~Meng, H.~Song and Y.~X.~Liu,
  %``Hybrid model approach for strange and multistrange hadrons in 2.76A TeV Pb+Pb collisions,''
  Phys.\ Rev.\ C {\bf 91}, no. 3, 034904 (2015).

%\cite{Zhu:2016puf}
\bibitem{Zhu:2016puf}
   X.~Zhu, Y.~Zhou, H.~Xu and H.~Song,
  %``Correlations of flow harmonics in 2.76A TeV Pb--Pb collisions,''
  Phys.\ Rev.\ C {\bf 95}, no. 4, 044902 (2017);  W.~Zhao, H.~j.~Xu and H.~Song,
  %``Collective flow in 2.76 A TeV and 5.02 A TeV Pb+Pb collisions,''
  arXiv:1703.10792 [nucl-th].

%\cite{Bernhard:2016tnd}
\bibitem{Bernhard:2016tnd}
  J.~E.~Bernhard, J.~S.~Moreland, S.~A.~Bass, J.~Liu and U.~Heinz,
  %``Applying Bayesian parameter estimation to relativistic heavy-ion collisions: simultaneous characterization of the initial state and quark-gluon plasma medium,''
  Phys.\ Rev.\ C {\bf 94}, no. 2, 024907 (2016).



%\cite{Song:2013gia}
\bibitem{Song:2013gia}
  H.~Song,
  %``Hydrodynamic modelling for relativistic heavy-ion collisions at RHIC and LHC,''
  Pramana {\bf 84}, 703 (2015); Nucl.\ Phys.\ A {\bf 904-905}, 114c (2013);  arXiv:1210.5778 [nucl-th].

%\cite{Luo:2011ts}
\bibitem{Luo:2011ts}
 X.~F.~Luo [STAR Collaboration],
%``Probing the QCD Critical Point with Higher Moments of Net-proton Multiplicity Distributions,''
 J.\ Phys.\ Conf.\ Ser.\  {\bf 316}, 012003 (2011).
		

%\cite{Luo:2011tp}
 \bibitem{Luo:2011tp}
  X.~Luo,
  %``Error Estimation for Moments Analysis in Heavy Ion Collision Experiment,''
  J.\ Phys.\ G {\bf 39}, 025008 (2012).

%\cite{Luo:2014rea}
\bibitem{Luo:2014rea}
  X.~Luo,
  %``Unified description of efficiency correction and error estimation for moments of conserved quantities in heavy-ion collisions,''
  Phys.\ Rev.\ C {\bf 91}, no. 3, 034907 (2015)
  Erratum: [Phys.\ Rev.\ C {\bf 94}, no. 5, 059901 (2016)].
 % doi:10.1103/PhysRevC.91.034907, 10.1103/PhysRevC.94.059901
 % [arXiv:1410.3914 [physics.data-an]].
  %%CITATION = doi:10.1103/PhysRevC.91.034907, 10.1103/PhysRevC.94.059901;%%
  %30 citations counted in INSPIRE as of 31 Jul 2017



\bibitem{Kapusta:2011gt}
  J.~I.~Kapusta, B.~Muller and M.~Stephanov,
  %``Relativistic Theory of Hydrodynamic Fluctuations with Applications to Heavy Ion Collisions,''
  Phys.\ Rev.\ C {\bf 85}, 054906 (2012).					
\end{thebibliography}
\end{document}